\documentclass[a4paper,12pt]{article}
\usepackage{amsfonts}
\usepackage{mathrsfs}
\usepackage{amsmath}
\usepackage{amssymb}
\usepackage[medium]{titlesec}
\usepackage{bm}
\usepackage{cite}
\usepackage[normalem]{ulem}
\usepackage{extarrows}
\usepackage{slashed}
\usepackage{isodateo}
\usepackage{graphicx}
\usepackage{xcolor}
\usepackage[bookmarksnumbered=true,bookmarksopen=true]{hyperref}
 \hypersetup{colorlinks,%
             linkcolor=[rgb]{0,0.2,0.6}, %
             citecolor=[rgb]{0,0.2,0.6}}
\usepackage[hmargin=.7in,vmargin=1.1in]{geometry}
\usepackage{indentfirst}
\newcommand{\FR}[2]{\displaystyle\frac{\,{#1}\,}{#2}}

\newcommand{\n}{\nonumber}

\graphicspath{{fig/}}

\def\bge{\begin{equation}}
\def\ede{\end{equation}}
\def\bga{\begin{aligned}}
\def\eda{\end{aligned}}
\def\bgp{\begin{pmatrix}}
\def\edp{\end{pmatrix}}
\def\bgs{\begin{subequations}}
\def\eds{\end{subequations}}
\newcommand{\order}[1]{\mathcal{O}({#1})}
\def\di{{\mathrm{d}}}

\def\mb{\mathbf}

\def\pd{\partial}

\def\la{\langle}\def\ra{\rangle}

\def\to{\rightarrow}

\def\al{\alpha}
\def\be{\beta}
\def\ga{\gamma}
\def\de{\delta}
\def\ep{\epsilon}
\def\ka{\kappa}

\def\rh{\rho}
\def\si{\sigma}

\def\hlbf{ }

\newcommand{\ob}[1]{\mkern 2mu \overline{\mkern -2mu #1 \mkern -2mu}\mkern 2mu}
\newcommand{\wt}[1]{\mkern 2mu \widetilde{\mkern -2mu #1 \mkern -2mu}\mkern 2mu}
\newcommand{\wh}[1]{\mkern 2mu \widehat{\mkern-2mu#1\mkern-2mu}\mkern 2mu}

\begin{document}
%\begin{fmffile}{ }
%\fancyhf{}
%\lhead{}
%\rhead{Begin on 2017/04/16, last updated on \TeXdate\today}
%\cfoot{\thepage}

\title{\vspace{-18mm}\Large\textbf{Induced Ellipticity for Inspiraling Binary Systems}}
\author{Lisa Randall$^a$\footnote{Email: randall@physics.harvard.edu} ~and~ Zhong-Zhi Xianyu$^{a,b}$\footnote{Email: xianyu@cmsa.fas.harvard.edu}\\[2mm]
\normalsize{$^a$~\emph{Department of Physics, Harvard University, 17 Oxford St., Cambridge, MA 02138, USA}}\\
\normalsize{$^{b}$~\emph{Center of Mathematical Sciences and Applications, Harvard University,}} \\
\normalsize{\emph{20 Garden St., Cambridge, MA 02138, USA}}}
\date{}
\maketitle

\begin{abstract}

 Although gravitational waves tend to erase eccentricity of an inspiraling binary system, ellipticity can be generated in the presence of surrounding matter. We present a semi-analytical method for understanding the eccentricity distribution of binary black holes in the presence of a supermassive black hole in a galactic center. Given a matter distribution, we show how to determine the resultant eccentricity analytically in the presence of both tidal forces and evaporation up to one cutoff and one matter-distribution-independent function, paving the way for understanding the environment of detected inspiraling black holes. We furthermore generalize Kozai-Lidov dynamics to situations  where perturbation theory breaks down for short time intervals, allowing  more general angular momentum exchange, such that eccentricity is generated even when  all bodies orbit in the same plane. 
\end{abstract} 

\section{Introduction}

Inspiraling binary systems formed by the merger of two compact objects such as black holes (BHs) or neutron stars (NSs) can generate characteristic and detectable gravitational wave (GW) signals. Such signals consist of three stages: inspiral, merger, and ringdown. The inspiral phase, generating a distinctive ``chirp signal'', is tractable within Newtonian dynamics at  leading order, and is potentially sensitive to ambient matter distributions. Given the precision with which these events can now be detected, it is worthwhile to ask whether we can hope to learn about the binary environment and not just the binaries themselves from such measurements. In particular, we explore in this paper how non-circular orbits can be generated in a dense local environment, ultimately focusing on the effect of orbiting near a large central BH.

It is well known that GWs tend to circularize orbits. Current LIGO searches therefore rely on templates that assume no ellipticity. However,   ellipticity can survive if there is some other source that drives an increase during inspiral or if the merger is detected soon after the binary was created, so that GWs could not erase the radiation. Such possibilities have been considered, including the Kozai-Lidov mechanism \cite{Kozai:1962zz, Lidov1976,Wen:2002km,VanLandingham:2016ccd,Hoang:2017fvh}, three-body systems \cite{OLeary:2005vqo,Gultekin:2005fd,Silsbee:2016djf}, and dense environments with many BHs created \cite{OLeary:2008myb}. All of these depend on details of the density distribution and binary formation, and can in principle be used to probe the assumed distributions.  

In this paper we consider eccentricity generation in the presence of a BH. We show how to understand the effect semi-analytically, showing how to incorporate both the nearby BH's tidal forces as well as evaporation when we assume a given density distribution. We give a simpler and more tractable understanding of the results of numerical integration of {\hlbf equations of perturbation} theory done by \cite{Antonini:2012ad}.

Coupled with previous work on active galactic nuclei (AGNs), for example, \cite{Bellovary:2015ifg,Bartos:2016dgn}, where it is argued that dynamical friction tends to help BHs migrate toward the galactic center,  elliptical orbits formed in the vicinity of a central BH could in principle be detectable by LIGO. Detailed interpretation might ultimately require full simulations, but we show which parameters the results depend on most heavily, paving the way for fitting results to BH mass and density profile more efficiently. We also demonstrate new dynamics that generalize the Kozai mechanism to more general exchange of angular momentum. This solution applies only when higher-order multipoles are included and we do not make the usual Kozai simplification of integrating over the mean anomalies of the two orbits. It is most effective at fairly eccentric large orbit so that the minimal distance distance between the binary and the tertiary body becomes small enough to violate fleetingly the condition for a perturbative expansion.

We proceed by showing how gravitational tidal forces will affect a GW signal in several different types of environments ranging from uniform to point-like matter distributions. To address the question of whether ellipticity can survive gravitational radiation, we consider several toy models. With increasing levels of complexity, we model the background by a uniform mass density, a nearby point mass, and an asymmetric cloud of massive objects. Our goal is to see how such backgrounds can affect the binary orbit and hence the chirp signal. In particular, we would like to study whether the binary system can be given an observably large eccentricity due to the perturbation of the background. We pay special attention to binary BHs in galactic centers, for which we show how to develop an analytical understanding of the Kozai-Lidov induced eccentricity distribution of such binaries when entering the LIGO window and how the resultant eccentricity can be determined in a simple way given one cutoff and one matter-distribution-independent function. For those already familiar with GWs, our chief results are in Sec.~\ref{sec_BgMatter}. Because we have in mind ultimately applying this type of analysis to both standard and nonstandard mass distributions, we include some review and toy cases before presenting the main results.

Before presenting the toy models, we briefly review in Sec.~\ref{sec_GWEmission} the standard Newtonian treatment of the inspiral phase and the well-known chirp signal for a circular orbit and also for generalized elliptical orbits. We pay special attention to the circularization process due to GW radiation.  In Sec.~\ref{sec_BgMatter}, we consider the perturbation to the binary system due to various kinds of massive backgrounds and also present a new fleetingly nonperturbative solution for a hierarchical triple. We collect some useful equations of classical perturbation theory applied to celestial mechanics in App.~\ref{app_PertKepler}, which is useful for our study of background perturbation in Sec.~\ref{sec_BgMatter}. In App.~\ref{App_Extra} we provide additional notes which are simply meant as review and introduction to readers not yet familiar with this field, including a brief review of GW generation in general relativity, some details about elliptical orbits, sample visualization of Kozai-Lidov solutions, and also some useful numbers regarding GW observations.

\section{GW Emission from an Inspiraling Binary System}
\label{sec_GWEmission}

In this section we review the basics of GW emission from an inspiraling  binary system, of which GW150914 detected by LIGO \cite{Abbott:2016blz} is a prime example. A GW-generating binary system typically undergoes three qualitatively distinct phases, namely, the inspiraling phase, the merger phase, and the ringdown phase. Among them, the inspiraling phase can be well understood using the post-Newtonian (PN) approximation to general relativity, and the leading-order effect from the mass quadrupole can already be captured by Newtonian physics. 

\subsection{Generation of GWs}
\label{sec_GGW}

We first consider the GW generation during the inspiral phase to leading order in the PN expansion, which serves as preparation for our study of ambient mass perturbations which will be presented in the next section. The presentation here mostly follows the treatment in \cite{GW}.

To begin, we consider a simplified case where the GWs are emitted from a binary system of two point masses $m_{1,2}$, with fixed circular orbit. Several simplified assumptions are made: 1) A circular orbit is assumed, with no eccentricity; 2) The orbit is fixed and thus the back-reaction from GWs is ignored; 3) The two objects, presumably BHs or NSs, are treated as point masses. We shall include the elliptical case and the back-reaction of GWs in the following, while we make the point-mass assumption throughout the section.

After adopting the three above assumptions, we are effectively dealing with a standard Kepler problem, which is in turn equivalent to a one-body problem with reduced mass $\mu=m_1m_2/(m_1+m_2)$ and orbit radius $R$. The motion can be described in the center-of-mass frame by
\bge
\label{motion}
  x =R\cos \omega t , ~~~y =R\sin \omega t , ~~~z =0,
\ede
where we have placed  the orbit in the $(x,y)$-plane for convenience. The orbital frequency $\omega$ is related to the radius $R$ by 
\bge
  \omega^2=\FR{G(m_1+m_2)}{R^3}.
\ede
The leading-order  GW (in transverse-traceless gauge) emitted by such a system is proportional to the mass quadrupole of the source, (We derive this formula in Appendix \ref{App_GW}.)
\bge
\label{gw}
  h_{ij}(t,\mb x)=\FR{1}{r}\FR{2G}{c^4}\Pi_{ij,k\ell}(\hat{\mb n})
   \ddot M^{k\ell}(t-r/c) ,
\ede
where $r$ is the distance between the source and the detector,\footnote{We assume the propagation occurs in flat space for convenience.} $\Pi_{ij,k\ell}(\hat{\mb n})$ is the projector onto transverse-traceless components along the direction $\hat{\mb n}=(\sin\theta\sin\psi,\sin\theta\cos\psi,\cos\theta)$, and the mass quadrupole\footnote{Strictly speaking, $M_{ij}$ is the second moment of mass, while the quadrupole moment should be defined as $Q_{ij}=M_{ij}-\de_{ij}M_{kk}$. But $M_{ij}$ and $Q_{ij}$ are interchangeable under the projector $\Pi_{ij,k\ell}$ since $\Pi_{ij,k\ell}\de_{k\ell}=0$.  } $M^{ij}$, defined in (\ref{MassQ}), is computed by $M^{ij}=\mu x^i x^j$ (we use $x^1=x$, $x^2=y$, $x^3=z$). Therefore, we have,
\bge
  \ddot M^{ab}=-2\mu\omega^2R^2\bgp \cos 2\omega t & \sin 2\omega t \\ \sin 2\omega t & -\cos 2\omega t \edp
\ede
Substituting the above expression into (\ref{gw}), or more explicitly, into its component form (\ref{hcomp}), and performing the time redefinition $\omega t\to \omega t-\psi-\pi/2$ to absorb the unobservable azimuthal angle $\phi$ and also an overall sign, we get to polarized components of $h_{ij}$,
\begin{align}
h_+(t)=&~A\Big(\FR{1+\cos^2\theta}{2}\Big)\cos 2\omega t ,\\
h_\times(t)=&~A\cos\theta\sin 2\omega t .
\end{align}
(See (\ref{hpolar}) for definition of polarizations $h_{+,\times}$.) The amplitude $A=4G^2\mu m/(c^4rR)$. The important point here is that the GW is monochromatic and the frequency of the GW $\omega_{GW}=2\omega$ is twice of the orbital frequency $\omega$.

It is customary to express the amplitude $A$ of the GWs in terms of the so-called chirp mass,  $M_c\equiv (m_1m_2)^{3/5}/(m_1+m_2)^{1/5}$, as,
\bge
  A=\FR{1}{2^{1/3}}\Big(\FR{R_c}{r}\Big)\Big(\FR{R_c}{\lambda}\Big)^{2/3},
\ede
where $R_c=2GM_c/c^2$ is the Schwarzschild radius of the chirp mass, and $\lambda=c/\omega_{GW}$ is the wavelength of the GWs.

We see that the GWs are circularly polarized when $\theta=0$ and linearly polarized in $h_+$ when $\theta=\pi/2$. In between, GWs should be elliptically polarized. Therefore, a measurement of polarization in principle reveals the inclination of the orbit plane relative to the line of sight.

In reality the orbit is not fixed because the radiated GWs drain energy from the source, so the radius of the orbit decreases with GW emission, and thus the frequency of the rotation increases, generating the characteristic chirp signal. To a good approximation, we can find the chirp behavior by equating the rate of energy lost in the binary system with the power of GW radiation. The total radiation power can be obtained from (\ref{totalP}), with the following result,
\bge
\label{power}
  P =\FR{32}{5}\FR{c^5}{G}\bigg(\FR{GM_c\omega_{GW}}{2c^3}\bigg)^{10/3},
\ede
while the total energy of the binary system is
\bge
  E=-\FR{Gm_1m_2}{2R}=-\bigg(\FR{G^2M_c^5\omega_{GW}^2}{32}\bigg)^{1/3}.
\ede
Equating $-\dot E$ with $P$, we get a differential equation for $\omega_{GW}(t)$, which can be easily integrated, giving
\bge
  f_{GW}(\tau)=\FR{1}{\pi}\bigg(\FR{5}{256}\FR{1}{\tau}\bigg)^{3/8}\bigg(\FR{GM_c}{c^3}\bigg)^{-5/8}\simeq 151\text{Hz}\bigg(\FR{M_\odot}{M_c}\bigg)^{5/8}\bigg(\FR{1\text{s}}{\tau}\bigg)^{3/8},
\ede
where $\tau=t_{\text{coal}}-t$ is the time before coalescence, and $f_{GW}=\omega_{GW}/(2\pi)$. Inversely,
\bge
  \tau\simeq 3.00\text{s}\bigg(\FR{M_\odot}{M_c}\bigg)^{5/3}\bigg(\FR{100\text{Hz}}{f_{GW}}\bigg)^{8/3}.
\ede

{\hlbf The generalization of above circular orbits to elliptical ones is well known, and we summarize the relevant results. More details can be found in App.~\ref{App_Ellip}. In essence, the motion of elliptical binaries is no longer uniform in time and thus the GWs emitted from an elliptical binary would develop harmonics with frequency integer multiples of the orbital frequency, $\omega_n=n\omega,~(n=1,2,\cdots)$. To characterize them, we introduce the standard orbital parameters, i.e. the semi-major axis $a$ and the eccentricity $0<e<1$, instead of the orbital radius $R$ used for circular orbits. For small eccentricity $e\ll 1$, the radiation power of GWs with frequency $\omega_n$ is given by,
\begin{align}
\label{Powern}
  P_n=&~\FR{32G^4\mu^2m^3}{5c^5R^5}F_n(e), \\
  F_1(e)=&~\FR{17}{192}e^2+\order{e^4},\\
  F_2(e)=&~1-8e^2+\order{e^4},\\
\label{Fn}
  F_n(e)=&~\FR{n^{2n}}{4^n\Gamma^2(n-1)}e^{2n-4}+\order{e^{2n-2}},~~~~(n\geq 3).
\end{align}
The result for a circular orbit is recovered when $e\to 0$, where only the $n=2$ component survives. On the other hand, for small $e\ll 1$, it is $F_3(e)\simeq 11.4e^2$ that gives the largest contribution other than $F_2(e)$. The signature of $e=0.1$, for example,  would be a harmonic with frequency $3/2$ times higher and amplitude one order of magnitude smaller than the base frequency $(n=2)$. This estimation can help us to understand the observability of small eccentricity.

On the other hand, for large eccentricity $e\sim 1$, the peak frequency of GWs can be much higher than the orbital frequency $\omega^2=Gm/a^3$. A good fit of the peak frequency as functions of $a$ and $e$ is given in \cite{Wen:2002km},
\bge
\label{GWpeak}
f_\text{peak}(a,e)=\FR{\omega_\text{peak}(a,e)}{2\pi}=\FR{\sqrt{Gm}}{\pi[a(1-e^2)]^{3/2}}(1+e)^{1.1954}.
\ede
For an order-of-magnitude estimation, it is enough to use the following approximate result,
\bge
\label{GWpeakApprox}
  f_\text{peak}(a,e)\simeq\FR{\sqrt{Gm}}{[a(1-e^2)]^{3/2}}.
\ede

The GWs carry both energy and angular momentum away from an elliptical binary and change its size and shape. The rate of the change is described by the well-known Peters' equations \cite{Peters:1964zz},
\begin{align}
\label{dadt}
\FR{\di a}{\di t}=&-\FR{64}{5}\FR{G^3\mu m^2}{c^5a^3}\FR{1}{(1-e^2)^{7/2}}\bigg(1+\FR{73}{24}e^2+\FR{37}{96}e^4\bigg),\\
\label{dedt}
\FR{\di e}{\di t}=&-\FR{304}{15}\FR{G^3\mu m^2}{c^5a^4}\FR{e}{(1-e^2)^{5/2}}\bigg(1+\FR{121}{304}e^2\bigg).
\end{align}
Eliminating $t$ and integrating the equation, we get,
\bge
  a(e)=a_0\FR{g(e)}{g(e_0)},
\ede
where the function $g(e)$ is,
\bge
\label{ge}
  g(e)=\FR{e^{12/19}}{1-e^2}\bigg(1+\FR{121}{304}e^2\bigg)^{870/2299}\simeq\left\{
  \begin{split}
    &e^{12/19}  &e\ll 1\\
    &\FR{1.1352}{1-e^2} & e\lesssim 1
  \end{split}
  \right.
\ede
Finally, we note that the orbital frequency $\omega^2=Gm/a^3$ of the elliptical binary system is determined by $a$ in a simple way, and that the time dependence of $a=a(t)$ as described in (\ref{dadt}) is dependent on the eccentricity. Therefore, in addition to the higher harmonics in the frequency spectrum, the eccentricity of the orbit can also be detected by accurately monitoring the time dependence of the base frequency $\omega_{GW}=2\omega$. Clearly the details of what will be observed depends on the detectors themselves. We leave the detailed question of observability to the LIGO team. For the purposes of our analysis, we use $e=0.1$ as a baseline for observability.

}

\subsection{Detectability of elliptical binaries}

LIGO  has designed its current templates for detecting circular binaries, though it is working to extend their analysis to elliptical orbits.  Both large and small ellipticities are potentially detectable \cite{Wen:2002km, Tai:2014bfa, Gondan:2017hbp}.  In highly elliptical systems binaries can merge when the minor axis becomes very small and ellipticity remains high. Small ellipticity  would lead to a distinctive waveform, which with the very sensitive LIGO detector should also be observable.

It is well known, however, that gravitational radiation in an isolated binary would circularize a binary orbit. The evolution of an  existing elliptical binary system would lead to loss of possible initial ellipticity through the emission of gravitational radiation.  Ultimately detectability depends on many details, but we can ask, for example, for the likelihood that an isolated binary system retains significant eccentricity ($e>0.1$) when its orbital radius falls into the LIGO window, namely, $\sim 1000$km (cf. Fig.\;\ref{fig_radius}).

\begin{figure}[tbph]
\centering
\includegraphics[width=0.45\textwidth]{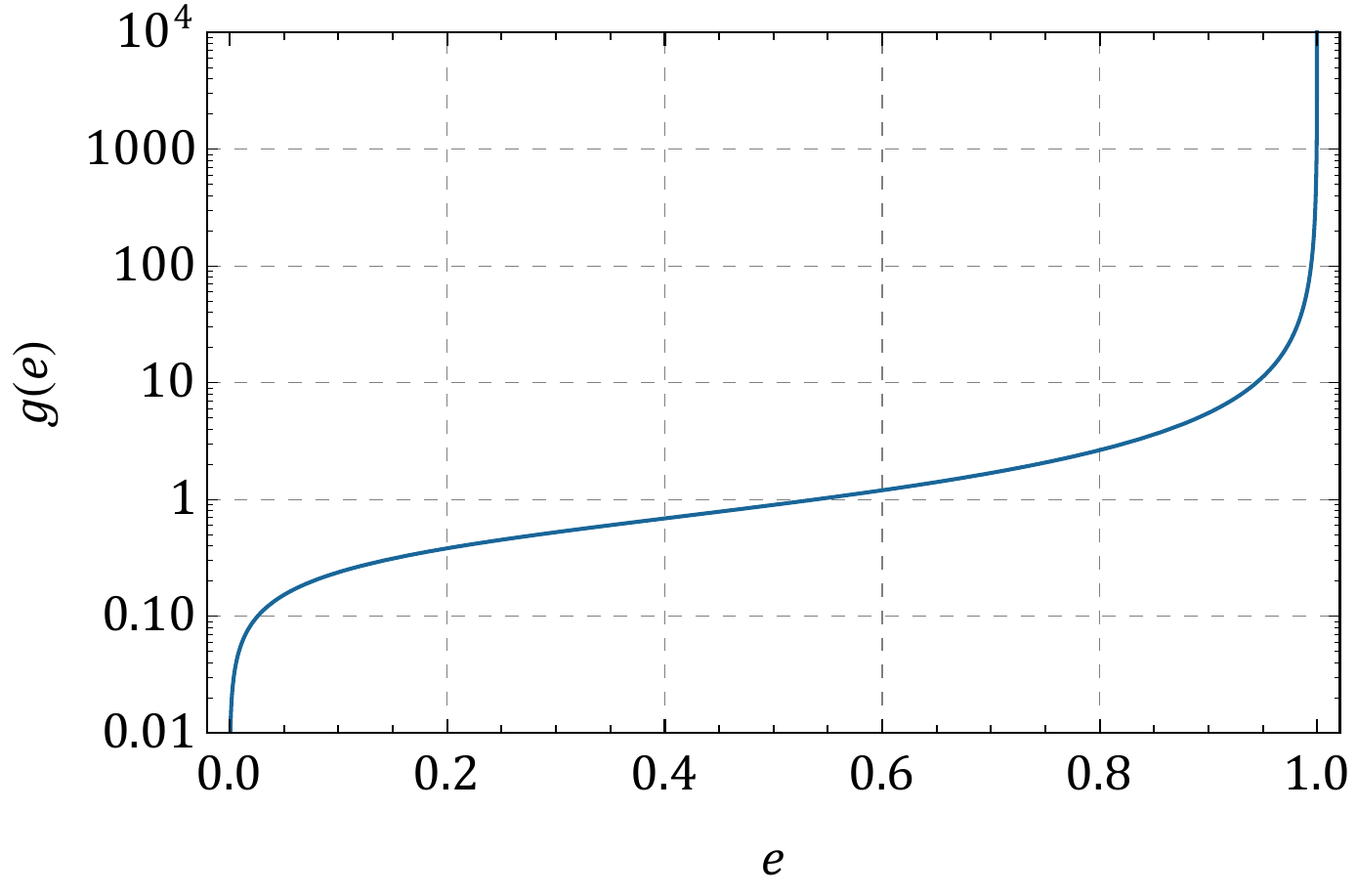}
\caption{The function $g=g(e)$ defined in (\ref{ge}).}
\label{fig_ge}
\end{figure}

A crucial ingredient of this estimation is the $g(e)$ function (\ref{ge}) derived in the previous subsection, which is plotted in Fig.\;\ref{fig_ge}.

For reasonable mean density $\rh$, the  initial separation of the binary masses, which we take roughly to be $0.1(M/\rh)^{1/3}$, would generally be much larger than the LIGO window ($\sim1000$km) by several orders of magnitudes. Therefore, in the absence of any ambient matter that can increase ellipticity, $g(e)$ tells us  that the eccentricity of the binary system at the moment of formation must be extremely close to 1 in order that the eccentricity remains reasonably large $(e> 0.1)$ when the orbital radius has reduced to $\sim1000$km. From Fig.\;\ref{fig_ge} we see that when the orbital radius decreases by an order of magnitude, $e$ reduces from 0.9 to 0.1, which means that for eccentricity greater than $0.1$ to survive, the eccentricity must be at least $0.9$ when the binary separation is $\sim10^4$km. Then, when $e>0.9$, we can use the approximate formula (\ref{ge}) for $e\lesssim 1$, which tells us that the required initial eccentricity $e_{\text{ini}}$ at the moment of formation should be,
\bge
  \FR{(M/\rh)^{1/3}}{10^4\text{km}}=\FR{1-0.9^2}{1-e_{\text{ini}}^2},
\ede
which gives,
\bge
\label{erho}
  1-e_{\text{ini}}^2\simeq \bigg(\FR{\rh}{10^{28}M_\odot/\text{pc}^3}\bigg)^{1/3}
\ede
The right-hand side of the above equation is an extremely small number for any reasonable density $\rh$. For example, if we take $\rh\simeq 10^{10}M_\odot/\text{pc}^3$, we will have $e_{\text{ini}}\simeq 1-10^{-6}$. Without any external driver of eccentricity, we would expect to observe only circular orbits in the LIGO window.

Previous work \cite{Wen:2002km} considers the possibility of inducing ellipticity through the passage of some other object.  The smallness of $1-e_\text{ini}^2$ is readily translated to the smallness of the impact parameter $b$, or the ``aiming angle'' $\theta$. If we assume that this happens early on in the binary formation or merging, the aiming angle must be exceptionally precise. This angle can be taken to be the ratio of the semi-minor axis $b$ and semi-major axis $a$, and therefore,
\bge
  \theta\simeq \FR{b}{a}=\sqrt{1-e^2_{\text{ini}}}.
\ede
On the other hand, the probability of aiming the target with such a small angle assuming random motion is roughly $\theta^2\simeq 1-e_\text{ini}^2$. This tells us that the probability of forming an elliptical orbit with LIGO-accessible eccentricity is roughly $10^{-6}$ for mean density $\rh=10^{10}M_\odot/\text{pc}^3$. Given the relation (\ref{erho}), we can write down an expression of the probability for general density, as,
\bge
  p\simeq 10^{-6}\times\bigg(\FR{\rh}{10^{10}M_\odot/\text{pc}^3}\bigg)^{1/3}.
\ede

It has been argued, however, that systems such as globular clusters can be dense enough that multi-body interactions are likely, giving rise to three-body effects \cite{OLeary:2005vqo,Gultekin:2005fd,Silsbee:2016djf}. A promising possibility might be the  Kozai-Lidov mechanism, in which angular momentum is transferred from the binary orbit with an accompanying change in inclination angle that together preserve angular momentum. These are certainly possible sources of eccentricity, though it is not clear that the expected rate will be sufficient for LIGO to observe \cite{Gondan:2017hbp}.

Below we consider a related scenario, in which the binary orbits a central BH. We will show that tidal forces can provide measurable ellipticity and consider semi-analytically how the Kozai-Lidov mechanism effects can be calculated. In a companion paper  we generalize the Kozai-Lidov mechanism to allow more general angular momentum transfer between the binary and the orbit around an external object.

\section{Effect of Background Matter}
\label{sec_BgMatter}

In this section,  we consider how an ambient medium could potentially affect observations at LIGO. In the first warm-up example we consider an idealized case of uniform ambient density and derive the  shift in the frequency of the chirp signal that would be generated.  We then consider the opposite non-uniform example of tidal effects from a nearby point mass. We present the well-known Kozai-Lidov mechanism that generates secular exchange between the inclination and the eccentricity of the orbit due to the tidal perturbation. We also illustrate another solution less relevant to SMBHs with fast and random generation of eccentricity at tertiary periastron due to the high eccentricity of the ``outer'' orbit. We subsequently consider how to evaluate the observability of eccentricity generated at the galactic center due to a central supermassive BH (SMBH). which is our major focus. We also present a third example in which we generalize the effect of a single nearby point mass to a cloud of massive objects, and show how the resulting eccentricity distribution is related to the density of the cloud, interpolating between uniform density and a single point mass.

\subsection{Uniform Ambient Density}

We first consider an idealized system with  the binary sitting in a nonzero uniform ambient mass density $\rh$. We assume for simplicity that the ambient mass is unaffected by  the binary. With this assumption, we can find a solution with an ansatz assuming a circular orbit around a common center $O$.   Let the distance from $m_{1,2}$ to $O$ be $R_{1,2}$. Then, using the ``shell theorem'' of Newtonian gravity, we can write down the equation of motion with this ansatz as,
\begin{align}
  m_1\omega^2R_1=&~\FR{Gm_1m_2}{R^2}+\FR{8}{3}\pi G\rh m_1 R_1,\\
  m_2\omega^2R_2=&~\FR{Gm_2m_1}{R^2}+\FR{8}{3}\pi G\rh m_2 R_2,
\end{align}
where $R=R_1+R_2$. From these two equations we see that the center $O$ is still given by the mass center of $m_1$ and $m_2$, but the frequency is now, 
\bge
  \omega^2=\FR{G(m_1+m_2)}{R^3}+\FR{8}{3}\pi G\rh.
\ede
We see that the circular motion is almost the same as in the previous section, except that the orbital frequency $\omega^2$ is shifted by a constant. In particular, the total energy of the system is again,
\bge
  E=-\FR{Gm_1m_2}{2R}=-\bigg(\FR{G^2M_c^5\wh\omega_{GW}^2}{32}\bigg)^{1/3},
\ede
but with the shifted frequency $\wh\omega_{GW}^2\equiv \omega_{GW}^2-\frac{32}{3}\pi G\rh$ that appears in the equation. On the other hand, the total radiation power $P$ is still given by (\ref{power}). The equation $-\dot E=P$ can again be integrated analytically, giving a relation between $\omega_{GW}$ and $t$ in terms of hypergeometric functions. If we assume that the ambient density is small, then we can expand the result and keep only the leading correction from nonzero $\rh$, 
\bge
  \tau\simeq 3.00\text{s}\bigg(\FR{M_\odot}{M_c}\bigg)^{5/3}\bigg(\FR{100\text{Hz}}{f_{GW}}\bigg)^{8/3}\bigg(1+\FR{16G\rh}{63\pi f_{GW}^2}\bigg).
\ede
Here $\sqrt{G\rh}$ has the dimension of frequency, and characterizes the frequency shift $\Delta f$ due to the background density. 

GW observations like LIGO or LISA should be capable of measuring the frequency shift $\Delta f$  by monitoring many cycles of chirp signals. Currently, for confirmed detection, the template used by GW observers should fit the signal in phase during the whole period of the inspiral phase. Therefore in principle detectors can be potentially sensitive to tiny frequency shifts or to the presence of additional modes. Of course the sensitivity depends on the nature of the templates and how signals are treated when they deviate slightly. We do a very crude estimate of the  sensitivity of the detector to this shift in frequency by asking that the correction $|\Delta N|$ to the number of cycles $N$ during the whole inspiral phase is smaller than 1. In reality, a match of wave patterns asks for more accuracy, $|\Delta N|\ll 1$. So the constraint $|\Delta N|<1$ should be feasible so long as it can be distinguished from background or noise. We emphasize that the effect of background matter would not just be a frequency shift, but a different time dependence of the signal.

The number of cycles is approximately,
\bge
  N\simeq 2.2\times 10^4\bigg(\FR{10\text{Hz}}{f_{\text{min}}}\bigg)^{5/3}\bigg(\FR{M_\odot}{M_c}\bigg)^{5/3},
\ede
where $f_{\text{min}}$ is the minimal frequency entering the detector. Therefore, let the shift of the frequency be $\Delta f$, then the corresponding shift in $N$ is roughly,
\bge
  \Delta N=-\FR{5}{3}N\FR{\Delta f}{f_{\text{min}}}.
\ede
Requiring $|\Delta N|> 1$  is therefore equivalent to requiring $\Delta f/f_{\text{min}}>N^{-1}$. For LIGO, this ratio can be $10^{-3}$, while for LISA, it can be $10^{-6}$. Therefore\footnote{Useful unit conversions: $\sqrt{GM_\odot/\text{pc}^3}\simeq 2\times 10^{-15}$Hz; $1\text{Hz}\simeq \sqrt{2\times 10^{29}GM_\odot/\text{pc}^3}$.}, for LIGO, we have $f_{\text{min}}\sim 10$Hz, and the detectable $\Delta f\gtrsim 10^{-3}f_{\text{min}}\sim 10^{-2}$Hz, which requires $\rh\sim 10^{25}M_\odot/\text{pc}^{3}$. While for LISA, we have $f_{\text{min}}\sim 10^{-4}$Hz, and $\Delta f\gtrsim 10^{-6}f_\text{min}\sim 10^{-10}$Hz, which gives $\rh\sim 10^{9}M_\odot/\text{pc}^3$. Clearly this will be a major challenge in either case. The more interesting situation is therefore the tidal effects that can be generated by a third body in reasonably close proximity, which is what we consider next.

\subsection{Tidal Effect from a Nearby Point Mass}

In this subsection we consider a nearby point mass $M$. For simplicity, we assume that the two members of the binary system have the same mass $m_1=m_2=m/2$. We consider only Newtonian dynamics in this subsection. In the following we will also include GW radiation. We also note that PN effects are small for our parameter range.  Let the binary separation be $R$ and the distance between the binary system and the nearby point mass be $L$. For sufficiently large $L$, we can treat the effect of the nearby point mass, or the tertiary body, as a perturbation. 

For perturbation theory to apply, we require that the tidal force generated by $M$ is smaller than the gravitational force between the two objects in the binary system. That is,
\begin{align}
  F_\text{tidal}\simeq GMmR/L^3 < G(m/2)^2/R^2\simeq F_\text{grav.},
\end{align}
which shows that the perturbation theory is good if $L>L_\text{pt}$, where
\bge
\label{Lpt}
  L_\text{pt}=\bigg(\FR{4MR^3}{m}\bigg)^{1/3},
\ede
or equivalently, if $(M/m)(R/L)^3<1/4$. When perturbation theory holds, the binary system still forms an elliptical orbit, but with secularly varying orbital parameters. We refer to the binary orbit as the ``small orbit'' or ``inner orbit''  and the orbit of the binary system  around the tertiary body along a generally elliptical path as the ``large orbit'' or ``outer orbit''. 

The analysis below is mostly fairly general but when we restrict our attention to a SMBH as the tertiary body later we will introduce the additional restriction that the orbit lies beyond $3 R_S$, where $R_S$ is the Schwarzschild radius associated with the tertiary mass $M$.

In fact, there are two distinct classes of solutions in which  eccentricity can be generated for the inner orbit. The first involves secular exchange between the inclination and the eccentricity of the inner orbit, which is known as the Kozai-Lidov mechanism. This class of solutions is characterized by slow variation in eccentricity but the value of eccentricity in each Kozai cycle can be very large, $e\sim 1$. 

There is also a second class of solutions in which the eccentricity of the outer orbit remains large, so that the perturbativity is violated for a short while whenever the binary reaches the periastron of the outer orbit. In this case, eccentricity can be generated for the inner orbit at each periastron of the outer orbit due to the violation of perturbation theory,  allowing angular momentum to be exchanged between the inner and outer orbits. This solution rarely generates sufficient eccentricity to survive to the LIGO window once the Schwarzschild length restriction is imposed, but it is interesting in its own right and thus we also present it in the following.

\subsubsection{Kozai-Lidov Solutions}

We begin by presenting a simplified analysis in which we assume that the outer orbit is circular with radius $L$, in which case the orbital frequency is $\Omega^2=G(M+m)/L^3$. We also assume that $L\gg L_\text{pt}$ with $L_\text{pt}$ as in (\ref{Lpt}) so that we can treat the effect of the tertiary body as a perturbation. 

We work in the center-of-mass frame of the binary system, in which the tertiary body $M$ has fixed distance $L$ from the the mass center of the binary. In this non-inertial frame, the effect of $M$ can be understood through an inertial force, namely the centrifugal force, plus gravitational forces. The combination of the centrifugal force and  gravity can be understood as a tidal perturbation, and such tidal perturbation can cause a secular modulation of orbital parameters of the binary, of which the eccentricity $e=e(t)$ is  our main focus. Note that we have chosen a nonrotating frame so a Coriolis force is absent. Were we to work in a rotating frame, the apparent orientation of the orbit would be different, but the dependence of $R$ and $e$ on $t$ would not.

\begin{figure}[tbph]
\centering
\includegraphics[width=0.5\textwidth]{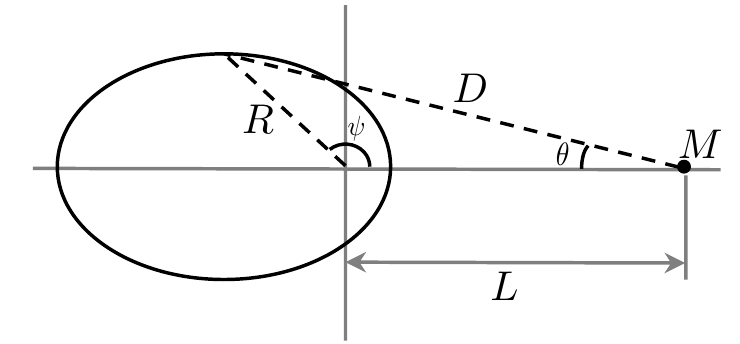}
\caption{A special configuration of the binary orbit with a nearby point mass $M$.}
\label{fig_orbit}
\end{figure}

Even without entering the details of a perturbative analysis, we can estimate the rate of change $\di e(t)/\di t$ due to this tidal perturbation. To this end, we note that the strength of the tidal force is,
\bge
  F_\text{tidal}\simeq \FR{GMmR}{L^3}.
\ede
For instance, in the simple configuration shown in Fig.~\ref{fig_orbit}, the tidal force from $M$ is,
\bge
\label{tidal}
  \mb F=-\mu\Omega^2L\hat{\mb x}+\FR{GM\mu}{D^2}(\hat{\mb x}\cos\theta-\hat{\mb y}\sin\theta)\simeq \FR{GM\mu}{L^3}\big(2\mb x-\mb y\big).
\ede
The rate of change of the eccentricity can be estimated by standard perturbation theory, as summarized in Appendix \ref{app_PertKepler}, with the following result,
\begin{equation}
  \omega_T\equiv\FR{\di e}{\di t}\bigg|_\text{tidal}\simeq \sqrt{\FR{GM^2R^3}{mL^6}}.
\end{equation}
To gain some intuition for this rate, we compare it to the inner orbital frequency $\omega^2=Gm/R^3$. We  see that $\omega_T/\omega=(\Omega/\omega)^2$. Perturbativity requires that $\Omega/\omega\ll 1$, so we see that the rate of change of the eccentricity is much slower than the angular velocities of both inner and outer orbits, and thus that the tidal perturbation can generate secular evolution of the eccentricity.

Going beyond the simple order-of-magnitude estimate requires a systematic treatment of perturbation theory. We use the explicit  equations for the mutual gravitational forces among the 3 bodies, and  perform a systematic expansion in $1/L$. The secular evolution of orbital parameters can then be solved by keeping the leading-order term in the expansion, known as the quadrupole interaction, and averaging over the periods of both inner and outer orbits. In this manner, the interaction can be viewed as an interaction between the inner orbit and the outer orbit. Notice that for the perturbative solution to hold, the rate of eccentricity development must be slower than any orbital frequency.

A careful analysis of this interaction shows that the quadratic interaction between the two orbits conserves the semi-major axes of both orbits, and hence the energies of the two orbits are also separately conserved. However, there can be secular exchange between the inclination and the eccentricity of the inner orbit, which is the well-known Kozai-Lidov mechanism\footnote{On the other hand, the next-to-leading-order term in the perturbation expansion of the total Hamiltonian, known as the octupole term, can sometimes be important, and leads to the so-called {\hlbf eccentric Kozai-Lidov mechanism \cite{Blaes:2002cs}.} The octupole term  is suppressed by the ratio of orbital radii for the small and large orbits. Since it is proportional to the mass difference of the binary, it vanishes exactly here.}. We shall not repeat the detailed analysis of this mechanism here and refer the readers to \cite{Lidov1976} for details. 
\subsubsection{Fleetingly Nonperturbative Solutions}

We now consider a separate category of solutions that allows more general angular momentum exchange and is relevant when the outer orbit has large eccentricity. Our solutions are distinct from the eccentric Kozai-Lidov solutions, which can also be relevant for large eccentricity of the outer orbit in that we will now consider a region where perturbation theory truly breaks down for a short time interval near the periastron of the outer orbit.   

The Kozai-Lidov mechanism works only for large inclinations, since the loss of eccentricity (at fixed major axis) with increased orbital eccentricity can then be compensated by decreased inclination angle. For small inclination, angular momentum can be conserved only by exchange with the outer orbit, but this does not happen at leading order in perturbation theory.

We see however that the binary orbit is not perfectly circular even for small inclination, as is clear from the first row of Fig.\;\ref{3dsolR}. {\hlbf However, direct few-body simulation shows that the binary orbit is not perfect circular even for small inclination.} This effect is not captured by the  Kozai-Lidov mechanism, since the generation of the eccentricity has a higher (or at least comparable) rate than the angular velocity of the large orbit and the solutions are derived after averaging both the inner and outer longitudes. The effect is entirely due to a breakdown of the perturbative expansion and can be seen only when  higher multiples allowing couplings between the inner and outer orbits contribute. \cite{Antonini:2012ad,Antognini:2013lpa} observed these fast oscillations about the Kozai solution numerically. Here we present a semi-analytical understanding that applies when the outer orbit has mildly large eccentricity. We note also that \cite{Antognini:2013lpa} did a similar but different analysis for highly inclined inner orbit. Note that in our case all even multipoles contribute and if we take different masses for the inner orbit, odd multipoles would as well. This is distinct from the eccentric Kozai-Lidov mechanism in which perturbation theory does not break down and only the octupole contributes.

\begin{figure}[tbph]
\centering
\parbox{0.47\textwidth}{\includegraphics[width=0.47\textwidth]{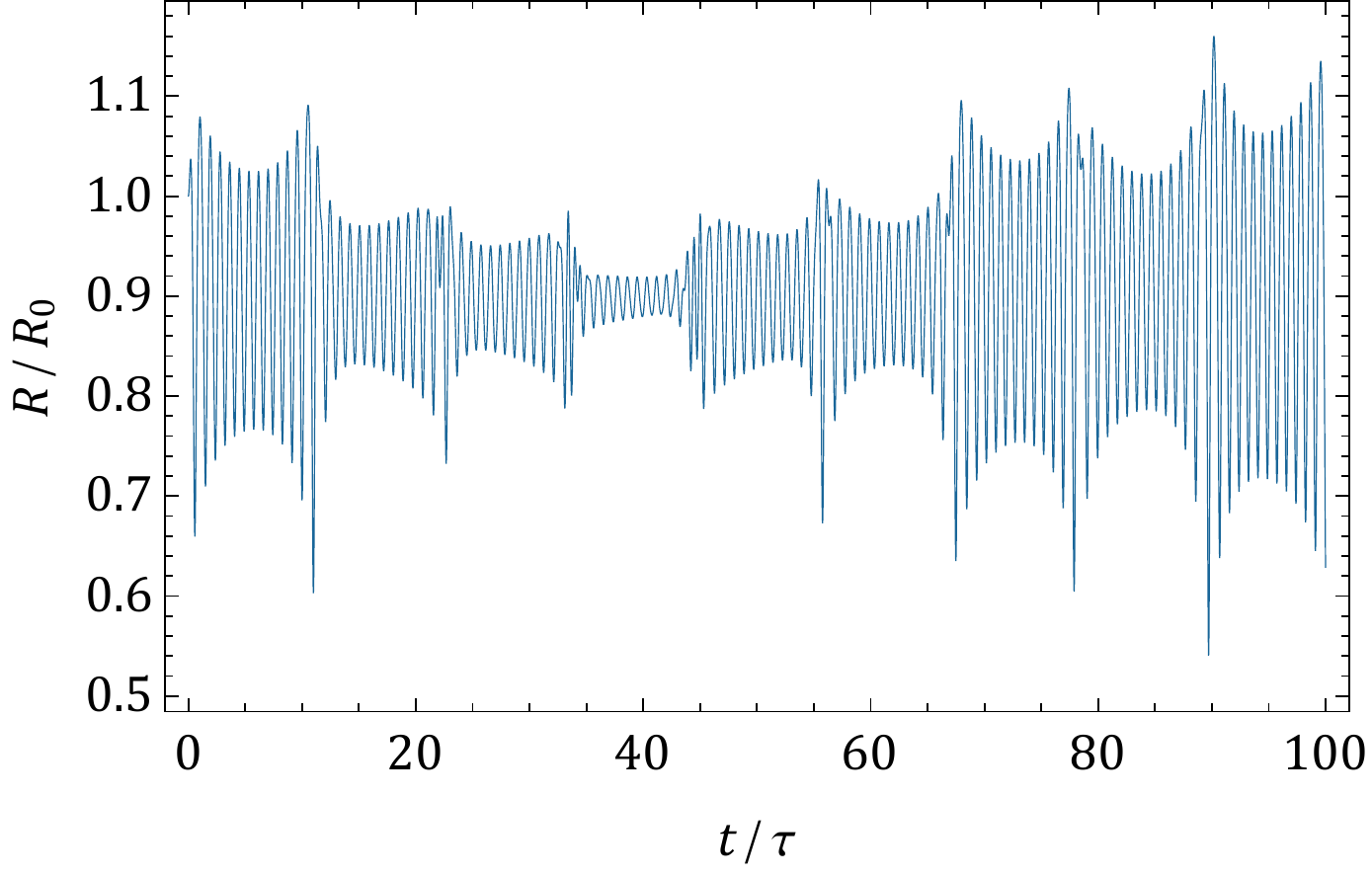}}
\parbox{0.44\textwidth}{\includegraphics[width=0.44\textwidth]{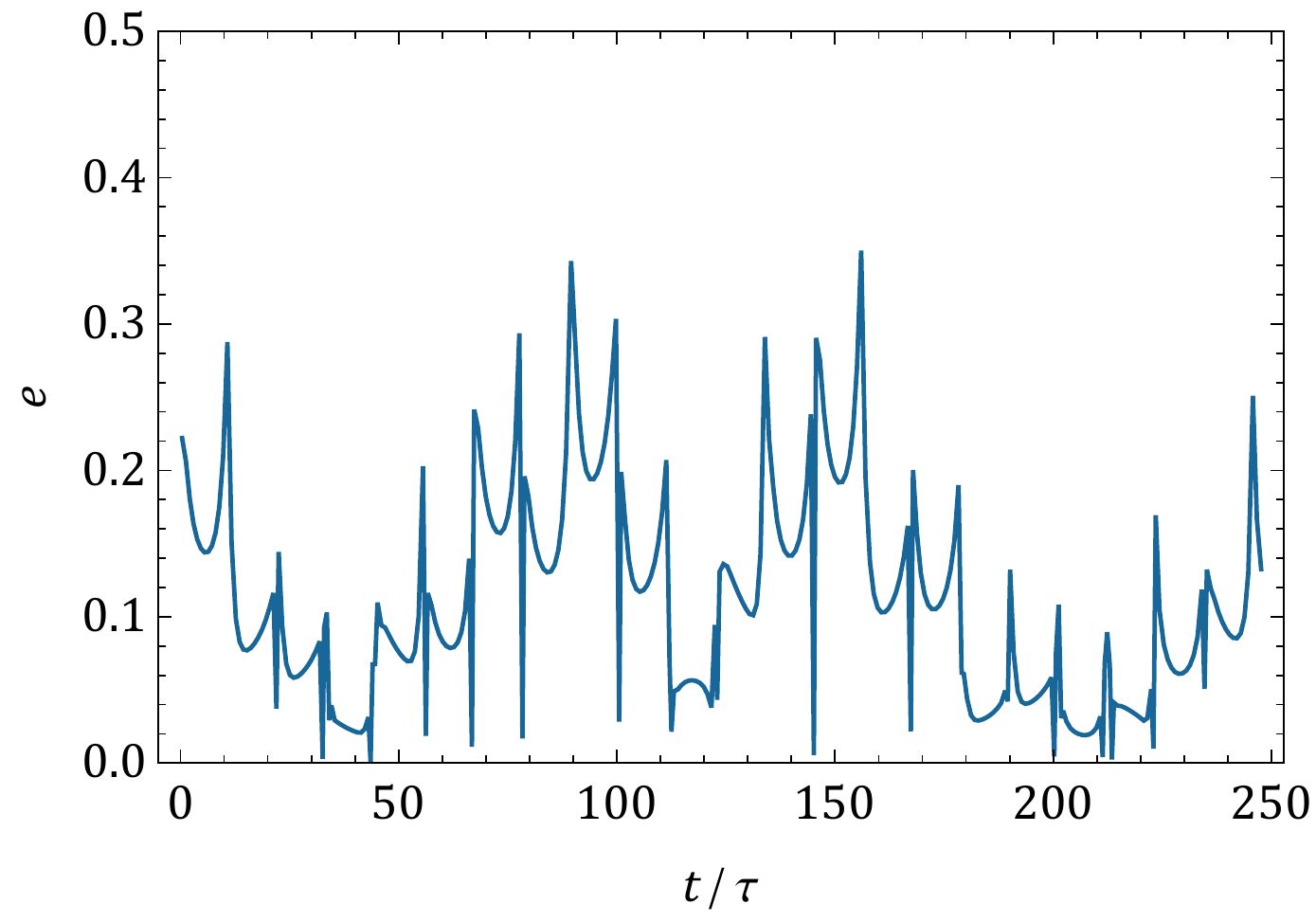}}
\caption{Sample 3d solution $a(t)$ and $e(t)$ with $e_0=0$, $e_{20}=0.45$, $(I_0,\ga_0,\vartheta_0)=(0^\circ,0^\circ,0^\circ)$, $M/m=10^3$, and $L_0/R_0=50$. The time $t$ is in the unit of inner orbital period $\tau=2\pi\omega^{-1}$.}
\label{3dsolR_e2}
\end{figure}

This rapid generation of eccentricity for a binary system becomes more significant when the large orbit is elliptical. In fact, if we again take the initial conditions $M/m=10^3$ and $L_0/R_0=50$ where $L_0$ represents the initial semi-major axis of the large orbit, and allow the large orbit to have an initial eccentricity $e_{20}=0.45$, we can then see from Fig.\;\ref{3dsolR_e2} that the binary orbit would in general remain elliptical at all times, with an eccentricity up to $\sim 0.3$. We note that the eccentricity generated for the binary system also depends on the initial values of inclination and the angle between the ascending node and the periastron. 

The rapid generation of mild to moderate eccentricity for the inner orbit is mainly due to the elliptical outer orbit. In the case that the outer orbit becomes very elliptical, perturbation theory can break down, but only for a short time interval when the binary system reaches the periastron of the outer orbit. The inner orbit can then be strongly disturbed, and the net effect is that the inner orbital parameters are changed very rapidly at the outer periastron. Because of the breakdown of perturbation theory, octupole and higher-order terms are relevant and permit angular momentum exchange between the inner and outer orbits. However, too large a perturbation will destabilize the orbit altogether. So although higher eccentricity is in principle possible, the fraction of stable high-ellipticity orbits will be small.

We can see that the outer orbit changes by plotting $L(t)$, which demonstrates that both the major axis and eccentricity of the outer orbit change. The variation in $L$ of order $0.03$ relative to the outer major axis is consistent with the ratio of outer and inner orbit of 30 and the mass ratio of 100.

This solution, however, does not allow for a buildup of eccentricity over many cycles as is true for the Kozai-Lidov solution. Although large eccentricity can be generated, it does not dominate as there is only a narrow window for large eccentricity before the solution becomes unbound. This means that unless the solution is generated fairly close to the third body where tidal forces are large, eccentricity will not survive to any measurable extent. 

This solution is not directly relevant in any case to a SMBH for which the binary cannot be closer than 3 times the Schwarzschild radius. However, this solution might be relevant when computing the overall eccentricity distribution and following detailed dynamics in a galaxy. We leave detailed investigation of these fleetingly nonperturbative solutions to future work.

\subsection{Tidal Sphere of Influence}

The above illustrates the generation of ellipticity. The critical remaining question is if and when such ellipticity could be observable in physical situations. Apart from the experimental issues, which we leave to the collaboration in the hopes of new templates sensitive to ellipticity, we need to demonstrate that the back reaction from GW emission, which causes rather fast circularization, does not eliminate the ellipticity too quickly for any sizable generation of eccentricity to occur before the merger.  A less important effect is PN precession of the binary orbit, on which we shall comment in the following. We note that closely related work was performed by L.~Wen in \cite{Wen:2002km}. {\hlbf See also \cite{Antonini:2013tea}.} Here we give a simpler way to estimate the size of the effect and will also ultimately include the consequences of evaporation.

We therefore compare the rate of change of eccentricity due to the tidal force and the rate due to GW emission. For GW radiation, it is known that
\bge
  \FR{\di e}{\di t}\bigg|_\text{GW}= -\FR{152}{15}\FR{G^3 m^3}{c^5a^4}\FR{e}{(1-e^2)^{5/2}}\bigg(1+\FR{121}{304}e^2\bigg),
\ede
On the other hand, the rate of secular evolution of the eccentricity due to the tidal force via the Kozai-Lidov mechanism can be calculated using classical perturbation theory at leading (quadrupole) order,
\bge
\label{dedttidal}
  \FR{\di e}{\di t}\bigg|_{\text{tidal}}=\FR{15}{16}\sqrt{\FR{GM^2a^3}{2mL^6}}e(1-e^2)^{1/2}(1-\cos^2 I)\sin2 \ga,
\ede
where $I$ is the inclination and $\ga$ is the angle from the ascending node to the periastron.

For our analytical estimate, it suffices to use the following approximate expressions,
\begin{align}
  &\FR{\di e}{\di t}\bigg|_\text{GW}\simeq -\omega_{C}\FR{e}{(1-e^2)^{5/2}}, 
  &&\omega_C\equiv\FR{10G^3m^3}{c^5a^4},\\
  \label{omegaT}
  &\FR{\di e}{\di t}\bigg|_\text{tidal}\simeq ~\omega_{T}e(1-e^2)^{1/2},
  &&\omega_T\equiv\sqrt{\FR{GM^2a^3}{2mL^6}}.
\end{align}
The factors involving $I$ and $\ga$ have been suppressed as they are in general of $\order{1}$. We shall shortly see that for small $I$ or small $\gamma$ the binaries in a galactic center will quickly evaporate before merger is possible. Therefore, taking these factors to be of $\order{1}$ is consistent.
 
For the same reason, we can also neglect the dynamical evolution of the argument of the periastron $\gamma$ so long as the precession of $\ga$ does not destroy the Kozai evolution. This is the case at Newtonian level where the precession acts coherently with Kozai cycles. The Newtonian evolution of $\gamma$ is again governed by perturbation theory as,
\bge
  \FR{\di\ga}{\di t}\bigg|_\text{tidal}\simeq\FR{\omega_T}{(1-e^2)^{1/2}},
\ede
where $\omega_T$ is given in (\ref{omegaT}). On the other hand, the PN precession has the rate,
\begin{align}
  &\FR{\di\ga}{\di t}\bigg|_\text{PN}\simeq \FR{\omega_\text{PN}}{1-e^2},
  &&\omega_\text{PN}\equiv\sqrt{\FR{G^3m^3}{c^4a^5}}.
\end{align}
When the PN correction $\omega_\text{PN}$ is comparable to the effect from the tidal perturbation, $\gamma$ undergoes precession that can act incoherently with the Kozai cycle and then destroy the Kozai resonance. This happens when $\omega_\text{PN}> \omega_T$, or more precisely, 
\bge
\label{omegaPNomegaT}
  \FR{{\di\ga}/{\di t}|_\text{PN}}{{\di\ga}/{\di t}|_\text{tidal}}\simeq \bigg(\FR{\omega}{\Omega}\bigg)^2\FR{R_m}{a}\FR{1}{(1-e^2)^{1/2}}>1,
\ede
where $R_m$ is the Schwarzschild radius associated with mass $m$. Here the factor $(\omega/\Omega)^2\gg 1$ as required by perturbation theory, but $R_m/a\ll 1$. So it is not immediately clear when the PN precession would destroy the Kozai resonance. We see from (\ref{omegaPNomegaT}) that the PN effect becomes more important for larger $L$, and thus the condition for PN-domination can be rewritten as a lower bound on $L$, i.e.,
\bge
  L>4200\text{AU}\bigg(\FR{a}{1\text{AU}}\bigg)^{1/3}\bigg(\FR{m}{M_\odot}\bigg)^{-2/3}\bigg(\FR{M}{10^6M_\odot}\bigg)^{1/3}(1-e^2)^{1/2}.
\ede
Therefore the PN precession is important for very distant binaries. However we shall see in the next subsection that the binaries at this distance would also end up with the evaporation rather than the merger, and thus we can ignore the PN precession for the binaries relevant to our analysis, which can merge before evaporation.

The tidal perturbation can increase or decrease the eccentricity of the binary orbit, so that the eccentricity remains $\order{1}$ averaged over a very long period of time, but GW radiation always erases eccentricity. So either large eccentricity is generated allowing the merger to happen quickly at small periastron or GW radiation decreases the eccentricity after it is generated.

We therefore compare three time scales: the orbital frequency $\omega^2\sim Gm/a^3$, the frequency of tidal modulation of the eccentricity $\omega_T$, and the rate of circularization due to GW radiation $\omega_C$. For the large mass $M$ scenario in this approximation, we can ignore the back reaction on the mass cloud due to the binary system and  treat the binary system as a  test mass, and the cloud as a fixed lattice or point source, ignoring dynamical friction.
It is straightforward to show that
\bge
\FR{\omega_T}{\omega}=\FR{M}{m}\bigg(\FR{a}{L}\bigg)^3=\bigg(\FR{\Omega}{\omega}\bigg)^2,
\ede
which is $\ll 1$ as required. On the other hand,
\bge
  \FR{\omega_T}{\omega_C}\simeq \FR{M}{m}\bigg(\FR{a}{R_m}\bigg)^{5/2}\bigg(\FR{a}{L}\bigg)^3,
\ede
where $R_m\equiv 2Gm/c^2$. This ratio is larger than $\omega_T/\omega$ by a factor of $(a/R_m)^{5/2}$. Therefore, it is possible that the eccentricity modulation is not fully erased by GW emission, in which case the residue eccentricity generated from the tidal perturbation remains, and in principle can serve as a probe of the nearby massive object.

We can associate a {\it sphere of influence} with each massive object in the cloud, by requiring that the eccentricity variation due to tidal perturbations would not be erased to below an observable threshold by the GW radiation when the binary system is within this sphere. The radius $L_i$ of this sphere is  determined by $|\di e/\di t|_\text{GW}\simeq |\di e/\di t|_\text{tidal}$, namely,
\bge
\label{Li}
  L_i=a\bigg(\FR{M}{m}\bigg)^{1/3}\bigg(\FR{a}{R_m}\bigg)^{5/6}(1-e^2).
\ede

Because we expect a given binary to have an approximately fixed orbital radius $L$ about the SMBH for circular large orbit, we reverse the inequality to determine  the semi-major axis $a_i$ when tidal forces no longer dominate (due to the reduction of $a$),
\bge
\label{ai}
  a_i=\bigg[L^6R_m^5\bigg(\FR{m}{M}\bigg)^2\FR{1}{(1-e_i^2)^6}\bigg]^{1/11},
\ede
which we shall call critical separation. 
 
Notice that this separation depends on eccentricity. In general, the relative rate of tidal force and GW changes with eccentricity so the binary moves in and out of the sphere of influence as it evolves.

What we ultimately require for the final result is the binary orbit size for each eccentricity when the tidal force will no longer dominate so that GW emission is solely responsible for the binary orbit's evolution and we can follow the inspiral phase analytically.  If we know the exact eccentricity distribution generated by tidal effects, we can integrate over eccentricity to find the final distribution. Without an exact calculation, there {\hlbf will be} some dependence on the eccentricity distribution in our final result. This distribution depends on the orbital parameters of the binary system and also its distance to the central BH but can be calculated at each $L$ without requiring the background density profile (which can be incorporated for any given model). Therefore, it is to be calculated only once for given inner and outer orbital parameters. An analytical understanding of this distribution will be considered in a future work.

We also know the semi major axis $a_\text{LIGO}$ as a function of final ellipticity and binary mass $m$  when the frequency of the binary enters the LIGO window, $f\geq f_\text{LIGO}\sim 10$Hz, as determined by [cf. (\ref{GWpeakApprox})],
\bge
\label{af}
  a_\text{LIGO}=\bigg(\FR{Gm}{f_\text{LIGO}^2}\bigg)^{1/3}\FR{1}{1-e^2}.
\ede

The regime where the necessary equalities hold is larger than the naive estimate based on assuming the critical separation $a_i$ in (\ref{ai}) lies within the LIGO window, since ellipticity  can be generated earlier corresponding to  larger binary separation so long as sufficient ellipticity survives the GW emission required to enter the LIGO range.
 We therefore do not require that $a_{i}\lesssim a_\text{LIGO}$. A reduction in semi-major axis $a$ after ellipticity is generated by tidal perturbations would not necessarily fully erase the observability of eccentricity.

For example, by examining the $g(e)$ function defined in (\ref{ge}), an elliptical orbit with semi-major axis $a$ and eccentricity $e\simeq 0.9$ would have eccentricity $e\simeq 0.1$ when the semi-major axis reduces by an order of magnitude to $0.1a$. Therefore, we introduce a parameter $\ka\equiv a_i/a_{\text{LIGO}}$, which represents the reduction of the orbital radius during the time between the onset of circularization and the time of entering the LIGO window. We call $\ka$ the reduction factor. 

\begin{figure}[tbph]
\centering
\includegraphics[width=0.55\textwidth]{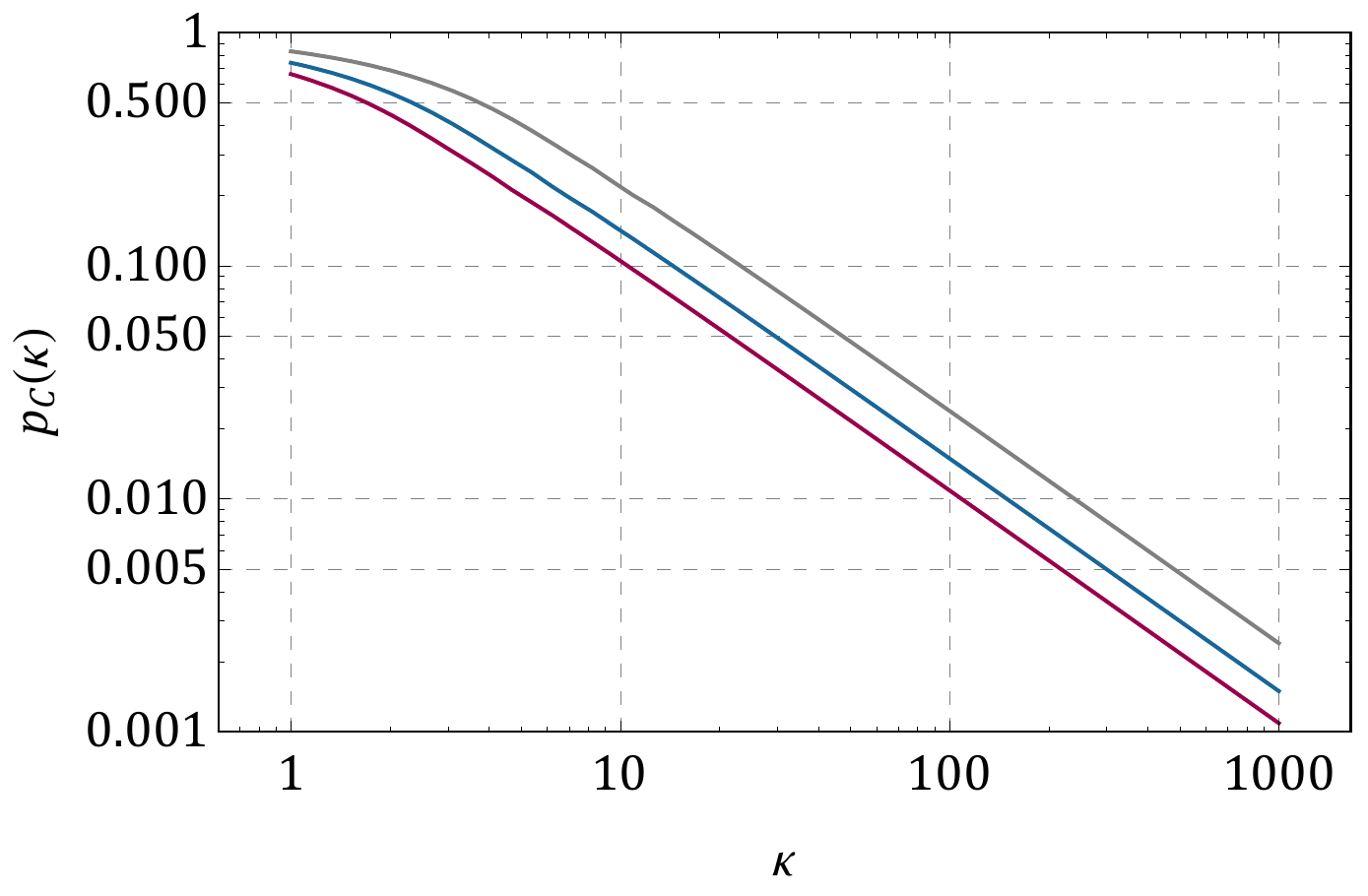}
\caption{The probability of finding a binary system with eccentricity larger than $e_\text{min}$ as a function of reduction factor $\ka$, assuming the eccentricity is democratically generated when circularization starts. The three curves from top to bottom correspond to $e_\text{min}=0.1,0.2,0.3$, respectively. }
\label{fig_pkappa}
\end{figure}

It is straightforward to see that the desired probability $p_C$ can be obtained in terms of the probability distribution $f (e)$ for finding an binary system with eccentricity $e$ by the time when GW radiation starts to significantly reduce the eccentricity.  We assume a power-law distribution $f (e)=(n+1)e^n$, then, 
\bge
\label{pC}
  p_{C}=\int_{g^{-1}(\ka g(e_\text{min}))}^1\di e\,f (e)=1-\Big[g^{-1}(\ka g(e_{\text{min}}))\Big]^{n+1},
\ede
where $g^{-1}$ is the inverse function of $g(e)$ given in (\ref{ge}). In Fig.\;\ref{fig_pkappa} we plot the above probability for $e_\text{min}=0.1,~0,2,~0.3$, and for a binary system of total mass $M_\odot$, assuming a uniform distribution $n=0$. 

\subsection{Tidal Perturbation from a Central BH}

We use the above  analysis  to study the tidal perturbation from a single SMBH. Assuming that the binary system stays at a fixed circular orbit around the SMBH with radius $L$, we can calculate the probability that the eccentricity of such a binary system is greater than 0.1 when entering the LIGO sensitivity window. This amounts to finding the appropriate value of $\ka$ for a given distance $L$. 

\begin{figure}[tbph]
\centering
\includegraphics[width=0.55\textwidth]{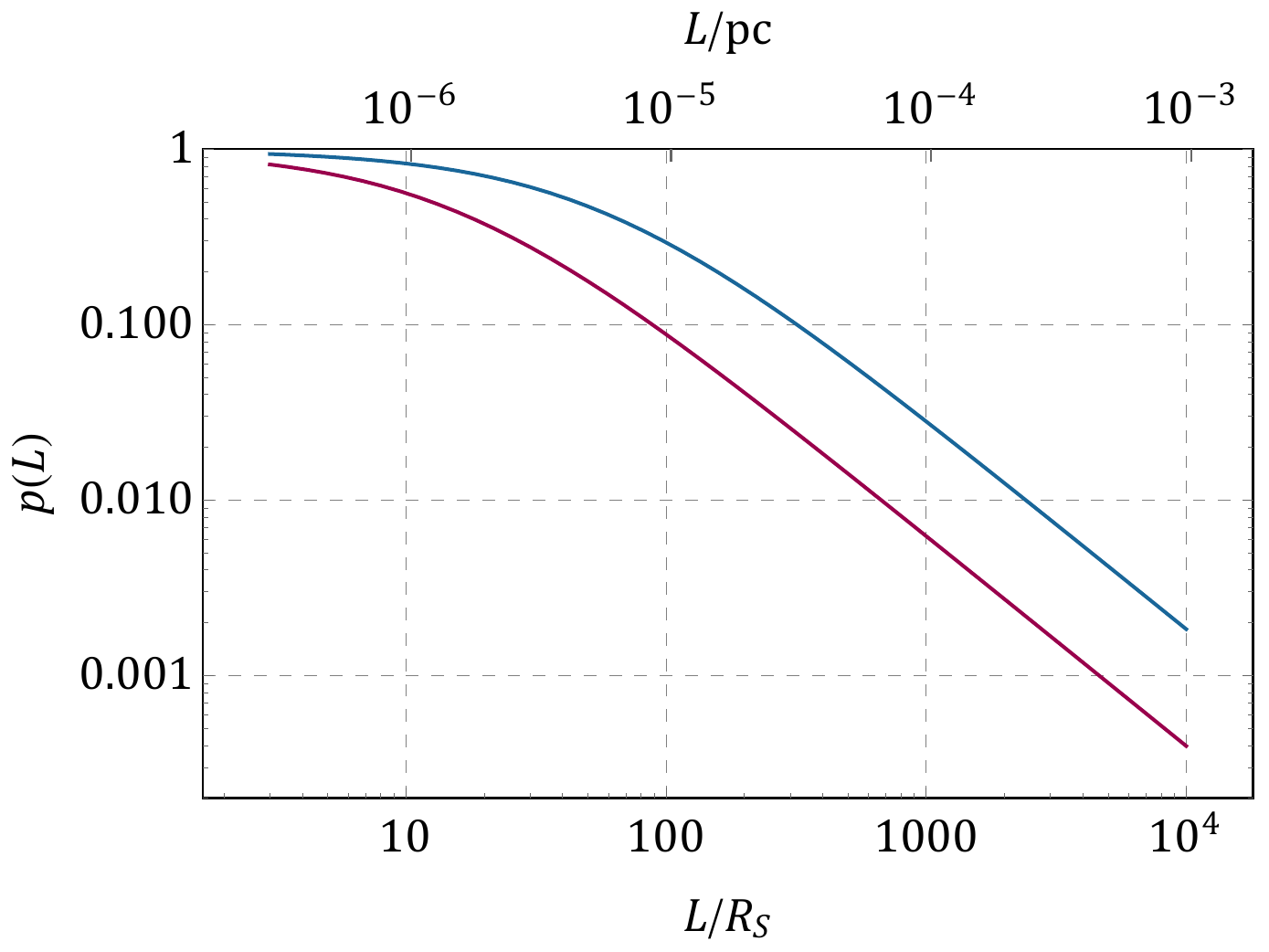}
\caption{The probability $p(L)$ that the binary system has eccentricity $e>0.1$ as a function of $L$, the distance between the binary system and the center of a SMBH of mass $10^6M_\odot$. The blue and orange curves correspond to the total mass of the binary system $m=M_\odot$ and $m=10M_\odot$, respectively. $R_S\simeq 3\times 10^6$km is the Schwarzschild radius of the SMBH. Note that the lower ends of the curves terminate at $L=3R_S$ which is the radius of the innermost stable circular orbit of a test particle outside a Schwarzschild BH.}
\label{fig_pL}
\end{figure}

To this end, we assume that the binary system sitting at a distance $L$ from the central SMBH  (which will be outside the SMBH horizon) has semi-major axis $a_i$ and eccentricity $e_i$ when  tidal effects cease to dominate, which then decreases to $a_f$ and $e_f$ when the binary enters the LIGO window. Here $a_i$ and $a_f$ are given by (\ref{ai}) and (\ref{af}), respectively, while the eccentricities are given by $a_i=g(e_i)$ and $a_f=g(e_f)$. As a result, we can find the probability (for a given initial eccentricity distribution) 
that the binary has eccentricity larger than a given value $e_f$ when entering the LIGO window to be,
 \bge
  p=1-e_i=1-\bigg\{\hat g^{-1}\bigg[\FR{(L^6R_m^5)^{1/11}}{(Gm/f_\text{LIGO}^2)^{1/3}}\bigg(\FR{m}{M}\bigg)^{2/11}(1-e_f^2)g(e_f)\bigg]\bigg\}^{n+1},
\ede
where $\hat g^{-1}$ denotes the inverse function of $\hat g(e)\equiv (1-e^2)^{6/11}g(e)$ and $g(e)$ is defined in (\ref{ge}), and the $(n+1)$'th power in the expression is from the power-law distribution of eccentricity $f (e)=(n+1)e^{n}$ as discussed above. For given masses and final eccentricity $e_f$, this probability is a function of distance $L$, which tells us how likely a BH binary can be observed with eccentricity larger than $e_f$ when sitting at a distance $L$ away from the central SMBH. We plot this function for different choices of $m$ in Fig.~\ref{fig_pL}, taking the uniform distribution $n=0$ as an example. 
 
This analysis was for a single binary system. The eccentricity distribution of BH binaries when entering LIGO's window depends on both the binary distribution in the galactic center and the binary   evaporation rate from stellar encounters. We do not model those ourselves, but instead follow the {\hlbf assumption of} Ref. \cite{Antonini:2012ad} in order to show we can closely reproduce the result they found for eccentricity distribution based on numerically integrating the equations of secular orbital evolutions plus GW radiation. 

To this end, we need the probability shown in Fig.~\ref{fig_pL}, as well as a profile of BH binary distributions in the galactic center. For the latter distribution, we take the one used in \cite{Antonini:2012ad}, i.e., the number density $n(r)\propto r^{-2}$. In addition, we introduce the inner cutoff of the binary distribution at the innermost stable circular orbit outside the central BH, i.e. three times of its Schwarzschild radius, and we introduce a cutoff on $a$ in accordance with perturbativity.

Key to the ultimate distribution is determining where the merger of the binary system is faster than its evaporation due to its interactions with background stars. This latter condition introduces an additional constraint on the binaries' inclination $I$ and  distance $L$. Essentially the dense background in the galactic center could disrupt the binary BHs and a binary system could ``evaporate'' due to the background disruption long before it merges. For generic eccentricity of the binaries, the merger time is rather long because the GW radiations are very weak for large binary separation. Therefore, most of the binaries in the galactic center would evaporate before they can merge.  

An important exception to this picture are the highly inclined binaries very close to the central BH. The high inclination translates to high eccentricity via the Kozai-Lidov mechanism, and the high eccentricities strongly boost the GW radiation so that the merging time of such highly inclined binaries can be strongly reduced. The closeness to the central BH enhances the tidal perturbation, making the Kozai-Lidov mechanism more efficient. As a result, some highly inclined binaries  close to the central BH  survive the background disruption and eventually reach the merger phase. In \cite{Antonini:2012ad} the evaporation and merger time are compared for a range of initial separations. Because the evaporation is a steeply falling function of $L$, we incorporate this as a numerically pre-determined cutoff $L_\text{cut}$ that we take from  \cite{Antonini:2012ad}. The ultimate eccentricity distribution is sensitive to this cutoff $L_\text{cut}$ beyond which binaries would evaporate too rapidly. The cutoff is a function of the initial cutoff on maximum binary separation as well as the full ellipticity distribution. Notice from \cite{Antonini:2012ad} that the result is a very steep function of $L$ ($a_2$ in those plots) so we can safely neglect binaries beyond the cutoff. 

\begin{figure}[tbph]
\centering
\includegraphics[width=0.45\textwidth]{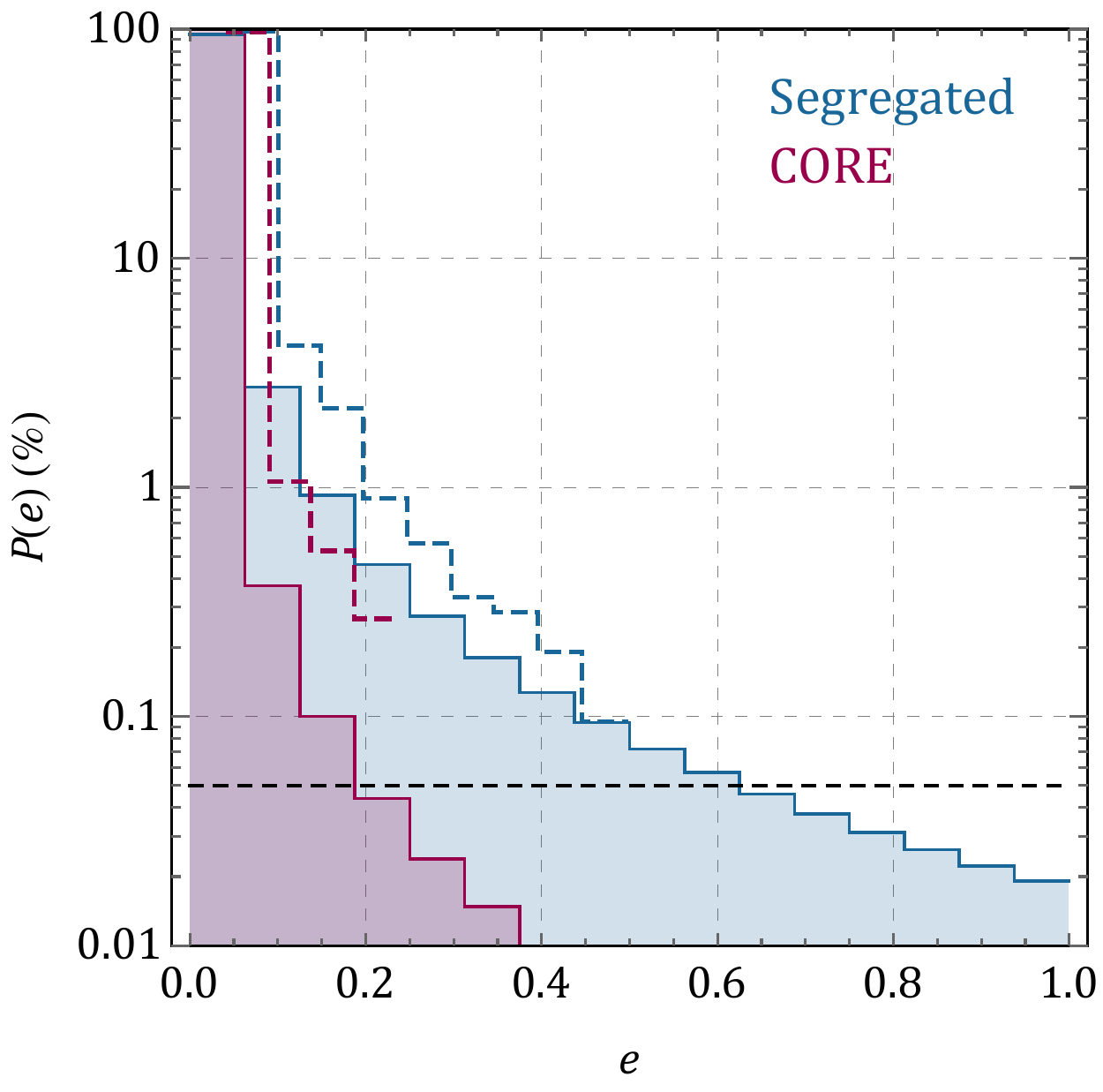}
\caption{The probability $p(e)$ of finding a binary BH by LIGO with eccentricity $e$ in a galactic nuclei. The blue and purple bins correspond to mass-segregated distribution and CORE distribution, respectively. The shaded distributions are computed in this paper assuming a distribution of initial eccentricity $f (e)\propto e^2$ when the circularization starts. The outer cutoffs due to evaporation time constraint are set at $L_\text{cut}\simeq 170$AU for mass-segregated model and $350$AU for CORE model. The dashed lines are corresponding results of \cite{Antonini:2012ad} from sampling the initial conditions and solving the combined equations of secular evolution and GW radiation. Note that \cite{Antonini:2012ad} has no data below $P(e)=0.05\%$, marked by the black dashed line.
}
\label{fig_pe}
\end{figure}

After taking account of  these constraints, we can plot the distribution of observed eccentricity for binaries in a galactic core, in Fig.~\ref{fig_pe}. It agrees well with the result of \cite{Antonini:2012ad} obtained via numerical integration of combined equations of secular evolution and the back reaction of GWs.

The results agree very well between our semi-analytical calculation and the more comprehensive numerical result of \cite{Antonini:2012ad}. This indicates the possibility of searching through effects of various stellar mass distributions without having to repeat the detailed analysis each time.  It also lets us zone in on what really can be potentially measured. Clearly the full matter density distribution will not be.

Our result depends on the initial distribution of the eccentricity at the beginning of the final circularization. We take a number of sample distributions in our plots but note that for a given distribution of binaries and central BH mass this distribution can be determined numerically, as was done in Ref. \cite{Antonini:2012ad}. Since the merger is more efficient for large $e$, we expect that this distribution $f (e)$ should be more peaked towards large $e$ compared with the initial distribution.   

\begin{figure}[tbph]
\centering
\includegraphics[width=0.55\textwidth]{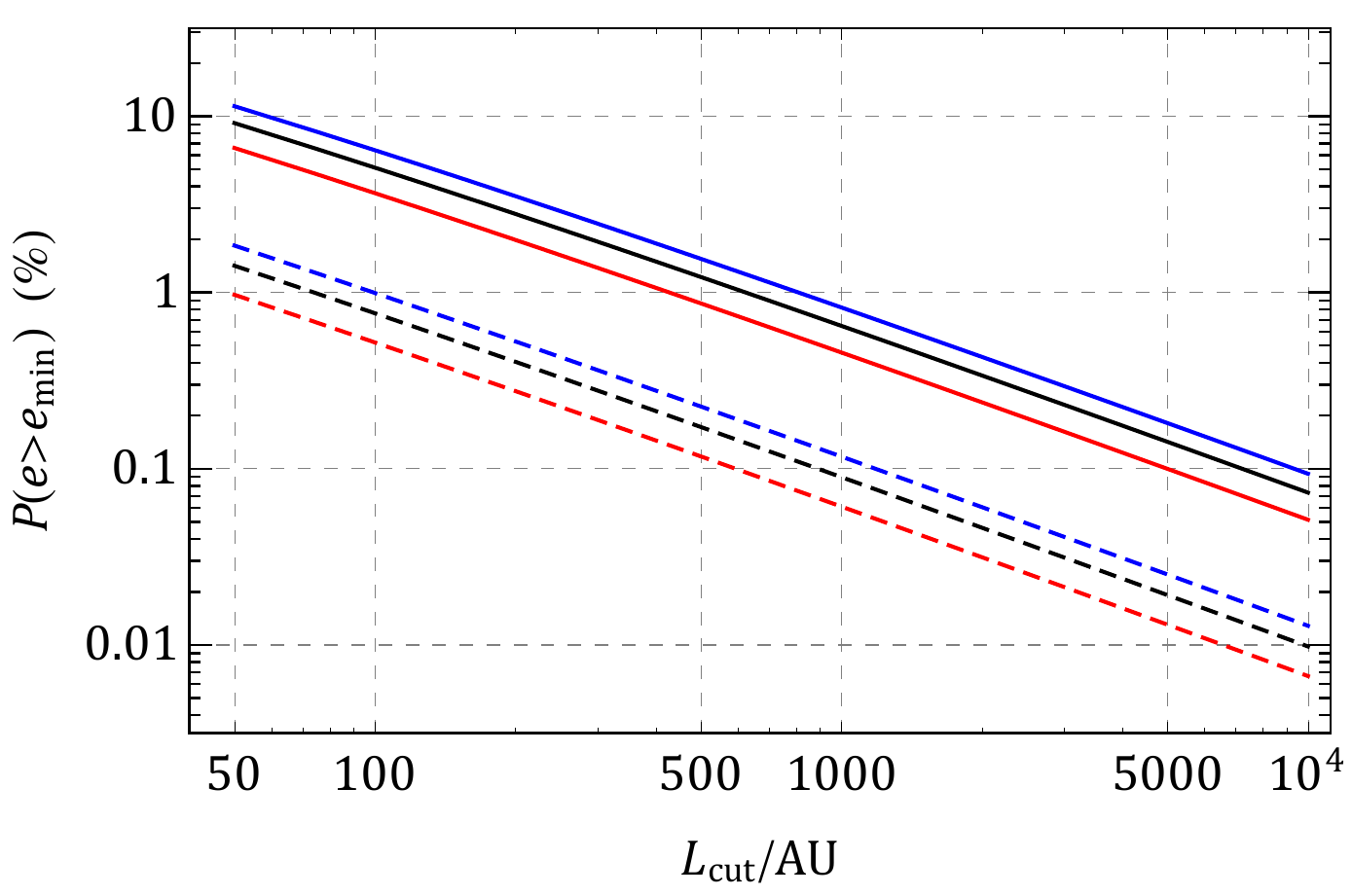}
\caption{The probability of observing a binary BH in galactic nuclei with eccentricity larger than a given value $e_\text{min}$ by LIGO, as a function of cutoff distance $L_\text{cut}$. The blue, black, red curves correspond to choosing distribution of eccentricity at the beginning of final circularization according to $f (e)=e^2$, $e$, $1$, respectively. The solid curves correspond to $e_\text{min} = 0.1$ and the dashed curves correspond to $e_\text{min} = 0.5$.}
\label{fig_PeLcut}
\end{figure}

\begin{figure}[tbph]
\centering
\includegraphics[width=0.55\textwidth]{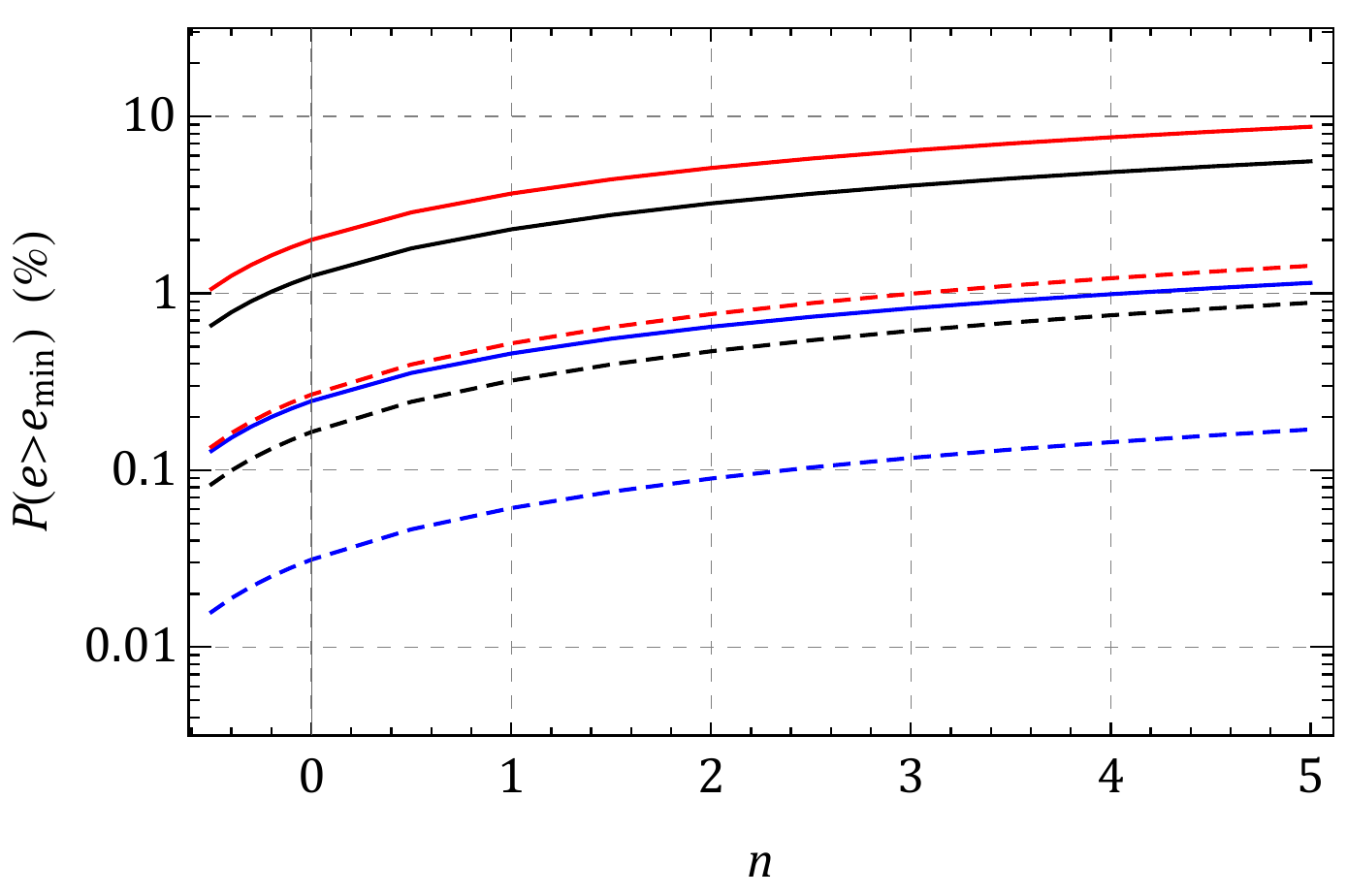}
\caption{The probability of observing a binary BH in galactic nuclei with eccentricity larger than a given value $e_\text{min}$ by LIGO, as a function of $n$ which is from $f (e)\propto e^n$. The red, black, blue curves correspond to choosing $L_\text{cut}=100,\;170,\;1000$, respectively. The solid curves correspond to $e_\text{min} = 0.1$ and the dashed curves correspond to $e_\text{min} = 0.5$.}
\label{fig_Pevsn}
\end{figure}

We plot our result as a function of cutoff $L_\text{cut}$ in Fig.\;\ref{fig_PeLcut} and as a function of eccentricity distribution $f (e)\propto e^n$ in Fig.\;\ref{fig_Pevsn}.

\begin{figure}[tbph]
\centering
\includegraphics[width=0.45\textwidth]{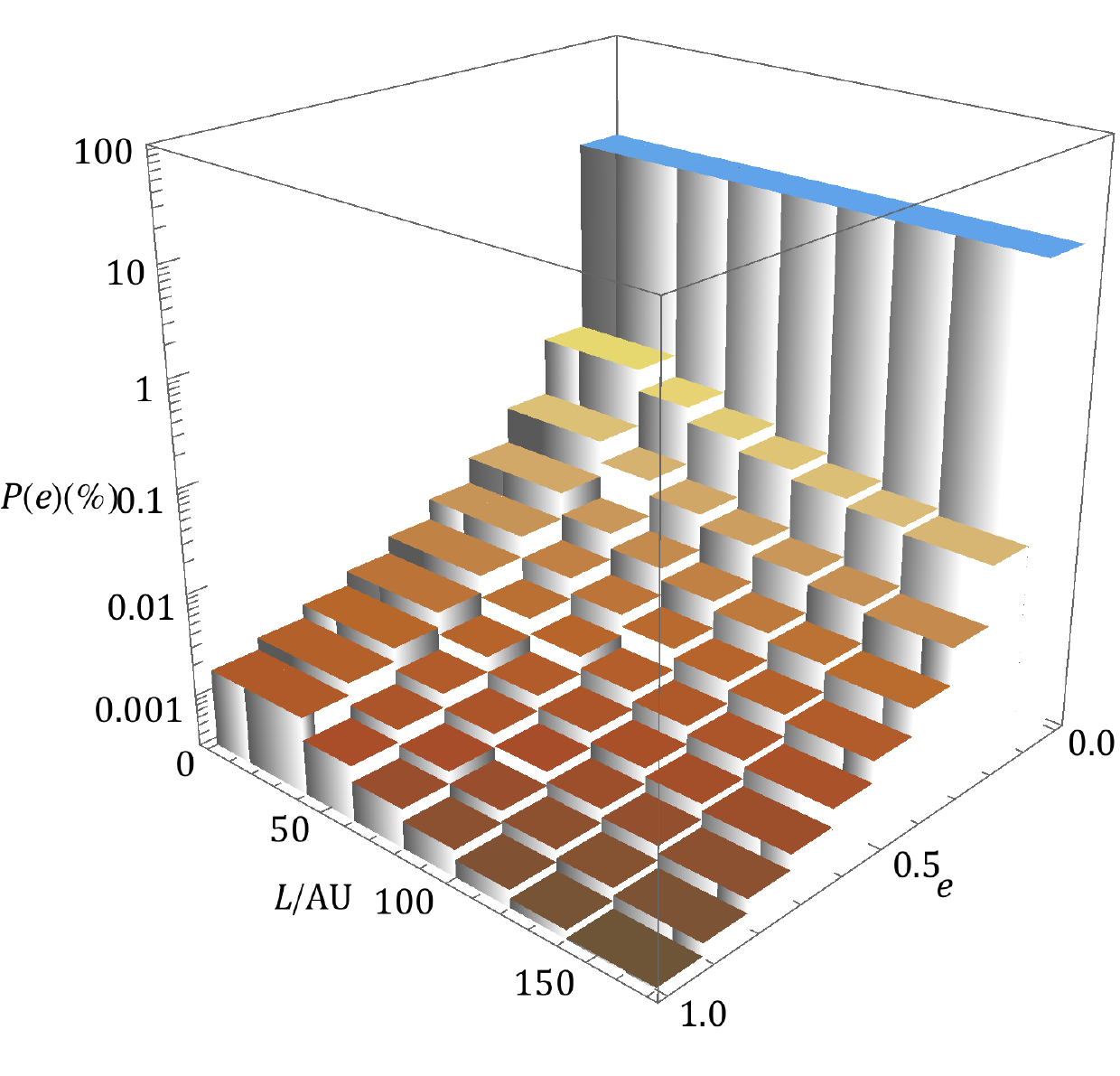}
\includegraphics[width=0.45\textwidth]{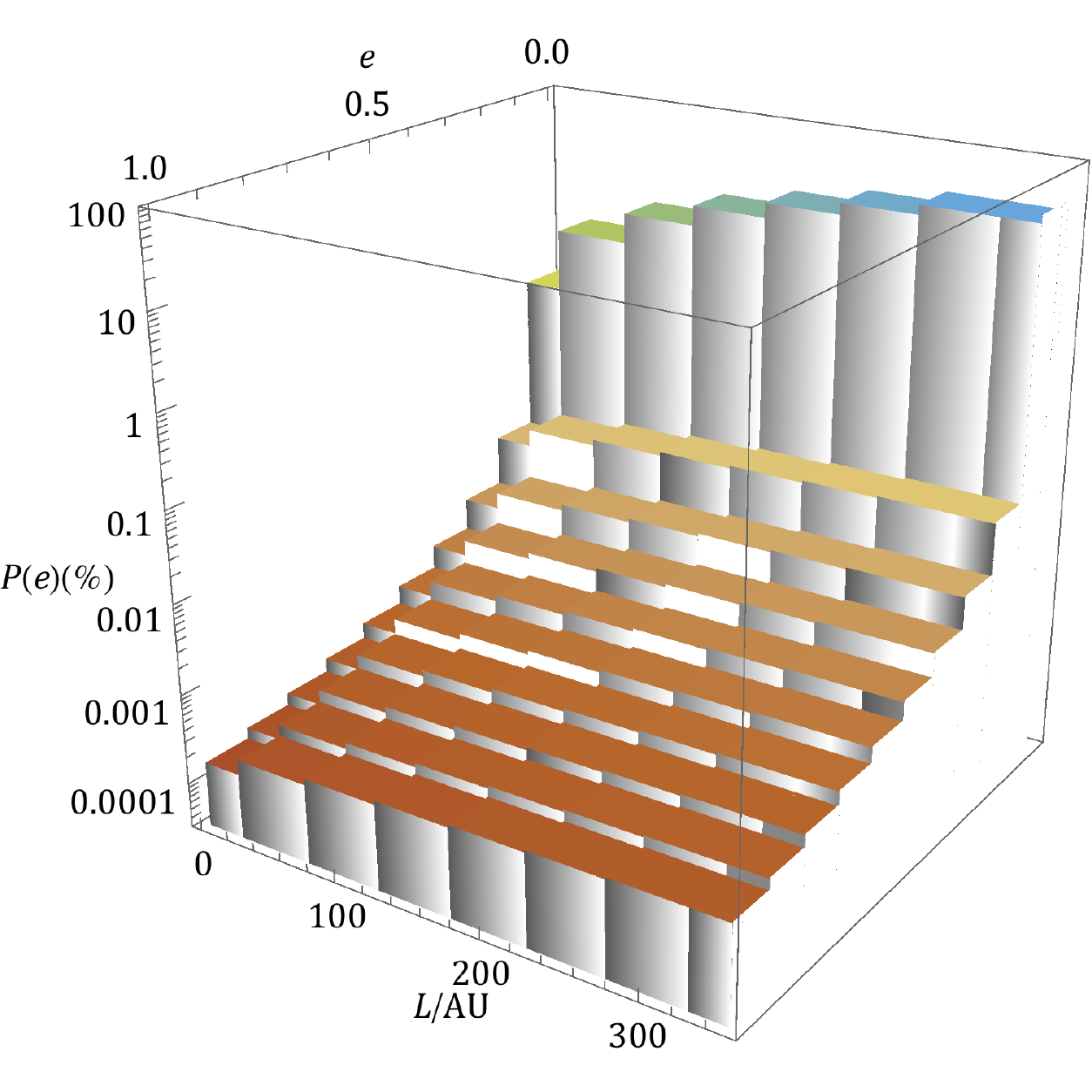}
\caption{The probability of observing a binary BH in galactic nuclei with eccentricity $e$ located at a distance $L$ from the central BH, assuming the mass segregated model (left) and CORE model (right) of binary distribution.}
\label{fig_Pe3d}
\end{figure}

Finally, it is also informative to present a breakdown of the eccentricity distribution according to the position of the binary BH near a central BH where it was generated, as shown in Fig,\;\ref{fig_Pe3d}.
We note there are several competing effects which this breakdown helps clarify. Evaporation is most effective in the denser central region but so is the Kozai-Lidov mechanism. On top of this, the outer region has greater volume so a distribution falling slower than $r^{-2}$ will be more weighted toward the outer region.

\subsection{Tidal Perturbation from a Cloud of Massive Objects}

Neither  uniform density nor a single nearby massive object captures the full range of possible matter distributions. In this section we briefly generalize our results by considering a binary system traversing a cloud of point masses. For simplicity we assume that the cloud is made from point masses $M$ spaced on a regular lattice and that the mean density is  $\rh$ when measured over the entire system.

A binary system will remain in an elliptical orbit so long as it stays close enough to an object that the tidal force dominates over the GWs, which defines  the  sphere  of influence. We estimate the probability that the binary system remains in an elliptical orbit by taking the ratio of the volume of a sphere of influence over the average volume occupied by a single massive object in the cloud. Note that the sphere of influence depends on the initial binary separation, which we parameterized via $\ka$. The result is,
\bge
  p_V(\text{LIGO})\simeq\FR{(8\pi/3)L_i^3}{d^3}\simeq\FR{\ka^{11/2}\rh}{9.9\times 10^{23}M_\odot/\text{pc}^3}\bigg(\FR{m}{M_\odot}\bigg)^{-5/3},
\ede
where $d\simeq(M/\rh)^{1/3}$ is the mean separation of cloud masses, and the subscript $V$ means that this part of the probability is obtained from comparison of the volume. We see that the result is quite sensitive to our choice of $\ka$. 
%For $\ka =10$, we have,
%\bge
 % p_V(\text{LIGO})=\FR{\rh}{3.1\times 10^{18}M_\odot/\text{pc}^3}\bigg(\FR{m}{M_\odot}\bigg)^{-5/3}.
%\ede
Interestingly, the result is independent of $M$, due to the inverse-cubic law of tidal force.  However, we emphasize that this $M$-independence holds only when $M$ is large enough to  consider only the tidal perturbation and to ignore dynamical friction. In particular, the uniform limit $M\to 0$ is not included in the formula derived here. Also keep in mind that $\rh$ is defined when averaged over the entire system, which is far from uniform when a particular mass $M$ dominating within each region of size  $d$. 

We can also estimate the result for LISA, 
\bge
  p_V(\text{LISA})=\FR{\ka^{11/2}\rh}{4.6\times 10^{9}M_\odot/\text{pc}^3}\bigg(\FR{m}{10^6M_\odot}\bigg)^{-5/3},
\ede
It is interesting to see how the result scales with $\ka$. We see that the sphere of influence and hence the relative volume scale as $\ka^{11/2}$ as $\ka$ goes up for generic $e$. 
Increased $\ka$ enlarges both the initial binary separation and hence the sphere of influence, allowing for larger probability for creating elliptical orbits. On the other hand,  larger $\ka$ means that the effect of the tidal perturbation will be erased earlier by GW radiation and thus the probability of a sizable eccentricity is reduced.  We also need to account for eccentricity dependence in both the tidal generation and  the circularization, which are especially important when $e\simeq 1$.  It is yet to be seen how the second effect, the longer circularization, scales as $\ka$ gets larger as well as the effect of large eccentricity $e\simeq 1$ by taking account of various $(1-e^2)$ factors in previous expressions.

\begin{figure}[tbph]
\centering
\includegraphics[width=0.55\textwidth]{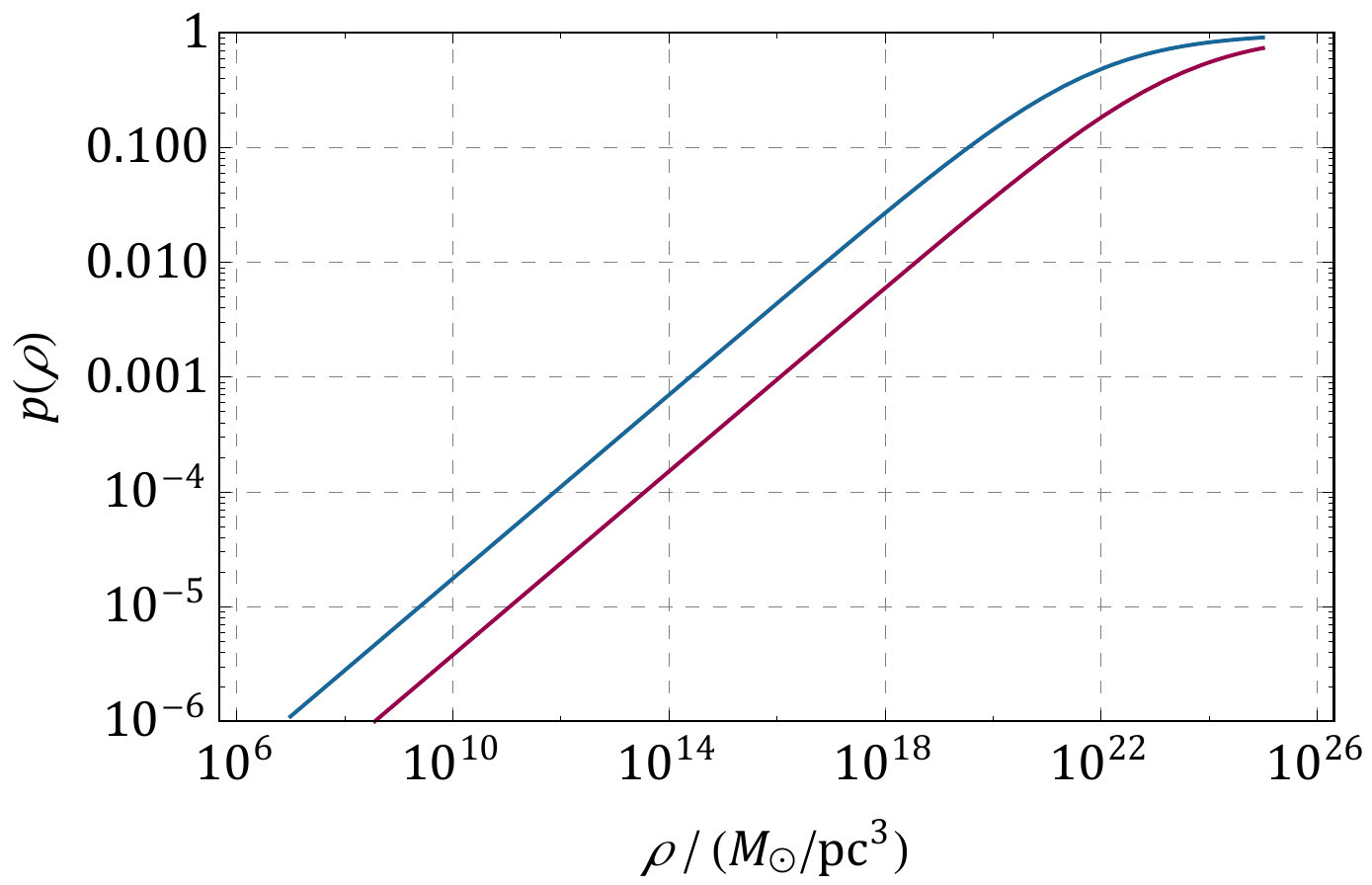}
\caption{The probability of observing a binary system with eccentricity larger than 0.1 as a function of background density $\rho$. The two curves from top to bottom correspond to the mass of binary system $m=M_\odot$ and $10M_\odot$, respectively.}
\label{fig_prho}
\end{figure}

To make this estimate, we need the probability distribution $f (e)$ for finding a binary system with eccentricity $e$ by the time when GW radiation starts to significantly reduce the eccentricity. This distribution can only be worked out by more careful analysis using classical perturbation theory. But for the moment, let us assume that the eccentricity follows a power-law distribution, i.e., $f (e)=(n+1)e^{n}$. Then, let us ask the question: Assuming the eccentricity is generated according to $f (e)$ just before the circularization starts, then what is the probability $p_C$ of finding a binary system with eccentricity $e$ larger than a given value $e_\text{min}$ when it enters the sensitivity window of LIGO, i.e., when its semi-major axis reduces to $a_\text{max}$. The subscript $C$ here refers to the circularization. 
\bge
  p=p_Vp_C.
\ede
However,  $p_V\propto R_i^3$ is suppressed by
$(1-e^2)^3$. When evaluated for initial eccentricity $e_\text{ini}=[g^{-1}(\ka g(e_\text{min}))]^{n+1}$, we see that $(1-e_\text{ini}^2)$ scales as $(n+1)/\kappa$ for large $\ka$ where $n$ is from the distribution $f (e)\propto e^n$. Consequently, $p_V$ actually scales as $\ka^{11/2}\cdot\ka^{-3}\propto \ka^{5/2}$ for large $\ka$. On the other hand, from (\ref{ge}) we see that $p_{C}\propto \ka^{-1}$ when $\ka$ is large. Therefore, we have $p=p_Vp_C\propto \ka^2$ for large $\ka$. Thus the optimal choice of $\ka$ is to make it as large as possible until the whole space is filled up by spheres of influence so that $p_V=1$ is saturated. We can then use this $\ka$, and find the probability from Fig.\;\ref{fig_pkappa}. The result is plotted in Fig.\;\ref{fig_prho}.
 
Of course the above analysis  takes account only  of the sphere of influence of  a single massive object near the binary. More generally, a cloud of objects can generate a tidal force, with some effects working in parallel to create eccentricity and others canceling out. We will not work out a detailed answer here, but simply note how we can understand the transition from a smooth distribution in which eccentricity will not be generated to one where we can focus on a single massive object.

Let the total mass of the binary system in question be $m$ and the orbital radius be $R$. Perturbation theory applies when $\rh\ll (m/R)^3$, which generally  holds for a reasonable background density $\rh$. Two physically distinct scenarios emerge: one with large $M$ so that the mean distance of cloud objects $d\sim(M/\rh)^{1/3}$ is much greater than $R$ and the other scenario with very small $M$ so that $d\ll R$. Basically, for given $\rh$, the mass $M$, and thus the mean distance $d$, characterize the smoothness of the cloud. When $M$ and $d$ are large, the cloud is very nonuniform,   and the analysis of the last subsection applies directly. An extreme case of this scenario is that we have only one object in the ``cloud'' with mass  $M\sim 10^{5\sim 7}M_\odot$, i.e., a SMBH as in the previous section. We shall consider this case in greater detail in the next subsection.
 
Next, we consider the  smoother scenario with small $M$ where the mean distance of cloud masses $d$ is comparable to or even much smaller than the orbital radius $R$ of the binary system. In such cases the back reaction on the cloud, and thus the dynamical friction, are not negligible. But let's just make a somewhat idealized analysis with the cloud fixed, and study how the tidal effect behaves for small cloud mass $M$. 

In this case, we can no longer use the approximate expression on the right-hand side of (\ref{tidal}), and a more general expression is needed. With the center-of-mass frame of the binary system with the mass center at the coordinate origin, we write the density of the cloud as a function of coordinates, $\rh=\rh(X,Y,Z)$. We also take the effective location of the binary system (see the plot of elliptical orbit in Sec. 3.2), then we can write the tidal force on the binary system as,
\bge
  \mb F=G\mu\int\di X\di Y\di Z\,\rh(X,Y,Z)\bigg(\FR{\mb D}{|\mb D|^3}-\FR{\mb L}{|\mb L|^3}\bigg),
\ede
where $\mb L=(X,Y,Z)$ and $\mb D=(X-x,Y-y,Z-z)$. This expression just follows the definition of the tidal force, which is the difference between the inertial force (the $\mb L$ term) and the gravitational force (the $\mb D$ term). For a cloud of massive objects, we have,
\bge
  \rh(X,Y,Z)=M\sum_i\de(X-X_i,Y-Y_i,Z-Z_i),
\ede
where the summation goes over all massive points. Then,
\bge
\label{tidalcloud}
  \mb F=GM\mu\sum_i\bigg(\FR{\mb D_i}{|\mb D_i|^3}-\FR{\mb L_i}{|\mb L_i|^3}\bigg),
\ede
where $\mb L_i$ and $\mb D_i$ are defined by the previous expressions with the argument replaced by $(X_i,Y_i,Z_i)$. It is now easy to see that, in the large $M$ limit, each massive point is sufficiently far away from the binary system, so that each pair of $\mb D_i$ and $\mb L_i$ is very close to one another. In this case, we are allowed to Taylor-expand $\mb D$ around $\mb L$, and the leading-order term is just proportional to the size of the binary system $\order{x,y,z}$, and the proportional constant is $GM\mu$ times some moments of the density function $\rh(X,Y,Z)$, which is of the same order as $GM\rh$, provided that $\rh$ is non-smooth enough. This is just the limit we have considered before, in particular, on the right-hand side of (\ref{tidal}).

On the other hand, we can now also understand the small $M$ limit with fixed $\rh$. In this case, the mean distance $d$ is comparable to or smaller than $R=\order{x,y,z}$ and thus we can no longer Taylor-expand $\mb D_i$ around $\mb L_i$. However, as $M$ goes down, the cloud gets smoother, and thus the moments of $\rh$ quickly decrease, as does the tidal perturbation. In the uniform limit $M\to 0$, the mean distance of the massive objects in the cloud goes to 0, and there is an approximate translation symmetry. Then  $\sum_i\mb D_i/|\mb D_i|^3$ is equal to $\sum_i \mb L_i/|\mb L_i|^3$ since the two are related to each other by translation. Consequently, there is no tidal perturbation at all. In particular, this implies that a circular orbit will remain circular for a uniform background, which agrees with our previous explicit solution.

\section{Conclusion}
We have recently  entered the era of GW detection. We have only just begun to ask what will be detectable and what these measurements might teach us. Clearly surprises are in store.

In this paper we addressed the question of how merging BHs can reflect their environment in the GW signal, and if so what would cause measurable effects. As in Refs. \cite{Antonini:2012ad}, we  have argued that a binary orbiting a super massive BH can give a signal that reveals an elliptical orbit or simply has a frequency shift from expectations. Current analyses are sensitive only to very small eccentricity or extremely large eccentricity if separate bursts are detected. We have seen that a large range of eccentricities occur for moderate values since even when the initial eccentricity is large GWs will reduce it. We leave the issue of generating new templates to the experimental collaborations.

In this paper, we have shown how to understand the magnitude of the eccentricity that can be generated and its distribution semi-analytically, so that we can ultimately scan through matter and BH profiles in AGNs or galactic centers to learn more about the environment in which eccentric BH mergers occur. We have shown that the resultant eccentricity distribution can be determined in a simple way up to the cutoff of the outer orbital radius and also the distribution of the eccentricity at the beginning of final circularization, which is independent of the matter distribution. We plan to study these two quantities more closely in a follow-up work. We also presented a fleetingly nonperturbative solution to the hierarchical triple system which shows fast generation of mildly large eccentricity even when the triple system is coplanar. In the future, it will be interesting to adapt our analysis to other compact binaries such as NS-NS pairs and NS-BH pairs.

Ultimately, as we make more observations and learn more about AGNs and galactic centers, we expect these results to help provide a very interesting probe that exploits the new window to the Universe provided by GW detection.

\paragraph{Acknowledgements.} We thank Prateek Agrawal, Imre Bartos, Luke Kelley, Frans Pretorius, Jakub Scholtz, and Kip Thorne for extremely useful conversations on several aspects of our paper. LR is supported by NSF grants PHY-0855591 and PHY-1216270. ZZX is supported in part by Center of Mathematical Sciences and Applications, Harvard University.

\begin{appendix}

\section{Perturbed Kepler Problem}
\label{app_PertKepler}

Classical perturbation theory is useful when studying tidal perturbations. 
For a binary system with elliptical orbit and reduced mass $\mu$, it is customary in perturbation theory to decompose the perturbation force in a special orthonormal basis $(\hat{\mb A},\hat{\mb Q},\hat{\mb J})$, where $\hat{\mb A}$ is chosen to be the unit vector in the direction of the Laplace-Runge-Lenz vector $\mb A$, and $\hat{\mb J}$ is the unit vector in the direction of angular momentum, and $\hat{\mb Q}$ is chosen to be orthogonal to both $\mb A$ and $\mb J$ and such that $(A,Q,J)$ form a right-handed orthonormal basis, as shown in the following plot.
\begin{center}
\includegraphics[width=7cm]{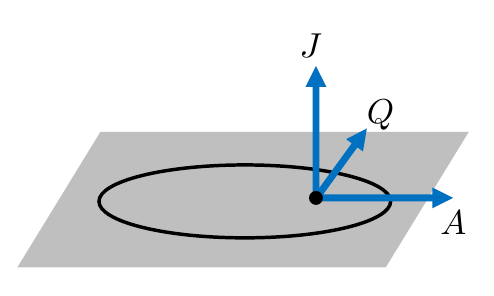}
\end{center}

Now we write the perturbation force in the following way,
\bge
  \Delta F/\mu= \al\hat{\mb A}+\be\hat{\mb Q}+\ga\hat{\mb J}.
\ede
Then the tidal force in (\ref{tidal}) can be written as,
\bge
  \al=\FR{3GM}{L^3},~~~~~~\be=-\FR{GM}{L^3}.
\ede
However, it should be noted that the vector $A$ itself is slowly varying under the tidal perturbation, so that the above expression holds only at one instance when $\mb A$ is in $x$ direction. But this fine detail does not affect our order-of-magnitude estimate. All we need to know is that $\al\sim\be\sim GM/L^3$.

From  perturbation theory, we know that the change of the Laplace-Runge-Lenz vector $A=|\mb A|$ due to the above perturbation is given by,
\bge
  \dot A=\be J+\FR{A+Gm\mu\cos\psi}{J}\mu R(\be\cos\psi-\al\sin\psi),
\ede
where $\psi$ is the azimuthal angle. Such a time dependence will generate an oscillation of eccentricity between 0 and 1 in time, with frequency given by the right-hand-side of the equation. To estimate this frequency, we use the fact that $\al\sim\be\sim GM/L^3$, and that the sinusoidal functions in $\psi$ can be approximated by a constant of $\order{1}$ after averaging over a quasi-period of orbital motion. For estimation, we can also take $\mu\sim m$ where $m$ is the total mass of the binary system, assuming that there is no large mass hierarchy between to two objects. Then, we note further that the magnitude of Laplace-Runge-Lenz vector is related to eccentricity via $A=Gm\mu e$, and that the angular momentum is given by$J^2=Gm\mu^2R$. Therefore,
\bge
  \dot e=\FR{\dot A}{Gm\mu}.
\ede
With all these ingredients, it is straightforward to derive (\ref{dedttidal}) quoted in the main text.

\section{Extra Notes}
\label{App_Extra}

\subsection{A Brief Review of GW Generation}
\label{App_GW}

In this appendix we review some basics of GW generation at the leading order from the mass quadrupole of the source.

\subsubsection{GWs from mass quadrupole} 

To derive the formula (\ref{gw}) for the GW emitted by a mass quadrupole at the leading order in PN expansion, we begin with the linearized Einstein equation in Lorentz gauge $\pd^\nu\ob{h}_{\mu\nu}=0$,
\bge
  \square \ob{h}_{\mu\nu}=-\FR{16\pi G}{c^4}T_{\mu\nu},
\ede
where $\ob{h}_{\mu\nu}\equiv h_{\mu\nu}-\frac{1}{2}\eta_{\mu\nu} h$, $h\equiv h_{\mu}{}^\mu$, and $h_{\mu\nu}\equiv g_{\mu\nu}-\eta_{\mu\nu}$. By construction we requires that $|h_{\mu\nu}|\ll 1$ holds over sufficiently large region in spacetime in some reference frame. All indices are to be raised and lowered by the flat metric $\eta_{\mu\nu}$.

The solution to the above equation with a given energy-momentum tensor $T_{\mu\nu}$ can be represented by the Green function $G(x-x')$ which is itself the solution to $\square G(x-x')=\de^{4}(x-x')$,
\bge
\label{Gsol}
  \ob h_{\mu\nu}=-\FR{16\pi G}{c^4}\int\di^4x'\,G(x-x')T_{\mu\nu}(x').
\ede
In radiation problem, we choose the retarded Green function,
\bge
\label{retGF}
  G(x-x')=-\FR{1}{4\pi|\mb x-\mb x'|}\de(x_\text{ret}^0-x'^0),
\ede
where $x'^0=ct'$, $x_\text{ret}^0\equiv ct_\text{ret}$, and $t_\text{ret}\equiv t-|\mb x-\mb x'|/c$.

Outside the source, we are free to choose the transverse-traceless (TT) gauge, which means that the ``graviton'' field $h_{\mu\nu}$ further satisfies $h^{0\mu}=0$, $h^i{}_i=0$, and $\pd^i h_{ij}=0$. For a given direction $\wh{\mb n}$ of propagation, it is convenient to introduce the projector $\Pi_{ij,k\ell}(\wh{\mb n})$ which projects a symmetric rank-2 tensor onto its TT component. Explicitly, this projector can be constructed as $\Pi_{ij,k\ell}(\wh{\mb n})=P_{ik}P_{j\ell}-\frac{1}{2}P_{ij}P_{k\ell} $ where $P_{ij}(\wh{\mb n})$ is itself a projector and is defined by $P_{ij}=\de_{ij}-n_i n_j$. Then, we can write down the expression for the graviton field in TT gauge outside the source, by substituting the retarded Green function (\ref{retGF}) into (\ref{Gsol}), and applying the TT projection,
\bge
  h_{ij}^{TT}=\FR{4G}{c^4}\Pi_{ij,k\ell}(\wh{\mb x})\int\di^3x'\,\FR{1}{|\mb x-\mb x'|}T^{k\ell}\Big(t-\FR{|\mb x-\mb x'|}{c},\mb x'\Big).
\ede
We are mostly interested in the far-field region with large $r\equiv |\mb x|$, where we can make the expansion $|\mb x-\mb x'|=r-\mb x'\cdot\wh{\mb x}+\order{d^2/r}$, and $d$ is the size of the system. Then at large $r$,
\bge
\label{hintT}
  h_{ij}^{TT}=\FR{4G}{c^4}\FR{1}{r}\Pi_{ij,k\ell}(\wh{\mb x})\int\di^3x'\,T^{k\ell}\Big(t-\FR{r}{c}+\FR{\mb x'\cdot\wh{\mb x}}{c},\mb x'\Big).
\ede 
At this stage we can apply the familiar multipole expansion to the energy-momentum tensor,
\bge
  T^{k\ell}\Big(t-\FR{r}{c}+\FR{\mb x'\cdot\wh{\mb x}}{c},\mb x'\Big)=T^{k\ell}(t-r/c,\mb x')+\FR{x'_i x^i}{c}\pd_{t} T^{k\ell}(t-r/c,\mb x')+\cdots.
\ede
For now, we keep the leading term only, which is the well-known quadrupole approximation.\footnote{For self-gravitating system like the inspiraling binary system we are considering in this note, to consistently include the higher multipole moments beyond quadrupole, it would be necessary to consider the  curved background. Therefore, a consistent PN expansion is needed beyond quadrupole order.} The name ``quadrupole'' derives from the fact that the integral in (\ref{hintT}) can be rewritten in terms of the mass quadrupole $M^{ij}$, defined by,
\bge
\label{MassQ}
  M^{ij}(t)=\FR{1}{c^2}\int\di^3x\,T^{00}(t,\mb x)x^i x^j.
\ede 
Using the covariant conservation of energy-momentum tensor $\pd_\mu T^{\mu\nu}=0$, it is trivial to prove the following identity,
\bge
  \ddot M^{ij}(t)=2\int\di^3x\,T^{ij}(t,\mb x),
\ede
by applying twice the integration-by-parts, and the boundary terms vanish because we assume that the source is nonzero only within a small region which is well contained in the integral region. Consequently, we reach the following formula for GW emission due to time variation of a mass quadrupole,
\bge
\label{hTT}
  h_{ij}^{TT}=\FR{1}{r}\FR{2G}{c^4}\Pi_{ij,k\ell}(\wh{\mb x})\ddot M^{k\ell}(t-r/c).
\ede
It is useful to rewrite the above equation in component form. Let's firstly consider the GW propagating in $\hat{\mb n}=\hat{\mb z}$ direction. Then the projector $P_{ij}=\de_{ij}-n_i n_j$ is simply given by diag$(1,1,0)$, and it is straightforward to obtain,
\bge
  \Pi_{ij,k\ell}M^{k\ell}=\bgp (\ddot M_{11}-\ddot M_{22})/2 & \ddot M_{12} & 0 \\ \ddot M_{21} && - (\ddot M_{11}-\ddot M_{22})/2 & 0 \\ 0 & 0 & 0\edp_{ij}.
\ede
In TT gauge, the GW $h_{ij}^{TT}$ can be parameterized as,
\bge
\label{hpolar}
  h_{ij}^{TT}=\bgp h_+ & h_\times & 0 \\ h_\times & -h_+ & 0 \\ 0 & 0 & 0 \edp_{ij}.
\ede
Comparing the two expressions, we get,
\bge
  h_+=\FR{1}{r}\FR{G}{c^4}(\ddot M_{11}-\ddot M_{22}),~~~~~~h_\times = \FR{2}{r}\FR{G}{c^4}\ddot M_{12}.
\ede
The result for GW $h_{ij}$ in a general direction $\hat{\mb n}=(\sin\theta\sin\phi,\sin\theta\cos\phi,\cos\theta)$ can be obtained by rotating the source $M_{ij}$ accordingly, namely using $M_{ij}'=R_{ik}^{-1}R_{j\ell}^{-1}M_{k\ell}$ as the new source, and substituting $M_{ij}'$ back into the above expression. Here the rotation matrix is given by,
\bge
  R=\bgp \cos\phi & \sin\phi & 0 \\ -\sin\phi & \cos\phi & 0 \\ 0 & 0 & 1 \edp
    \bgp 1 & 0 & 0 \\ 0 & \cos\theta & \sin\theta \\ 0 & -\sin\theta & \cos\theta \edp.
\ede
Then,
\bgs
\label{hcomp}
\begin{align}
  h_+=&~\FR{1}{r}\FR{G}{c^4}(\ddot M'_{11}-\ddot M'_{22})\n\\
     =&~\FR{1}{r}\FR{G}{c^4}\Big[\ddot M_{11}(\cos^2\phi-\sin^2\phi\cos^2\theta)+\ddot M_{22}(\sin^2\phi-\cos^2\phi\cos^2\theta)-\ddot M_{33}\sin^2\theta\n\\
      &~~~~~~~~~-\ddot M_{12}\sin 2\phi(1+\cos^2\theta)+\ddot M_{13}\sin\phi\sin2\theta+\ddot M_{23}\cos\phi\sin2\theta\Big],\\
  h_\times=&~\FR{2}{r}\FR{G}{c^4}\ddot M'_{12}\n\\
          =&~\FR{1}{r}\FR{G}{c^4}\Big[(\ddot M_{11}-\ddot M_{22})\sin 2\phi\cos\theta+2\ddot M_{12}\cos2\phi\cos\theta\n\\
           &~~~~~~~~~-2\ddot M_{13}\cos\phi\sin\theta+2\ddot M_{23}\sin\phi\sin\theta\Big].
\end{align}
\eds

\subsubsection{Energy and angular momentum of the quadrupole radiation}

In linearized theory, it is possible to write down an expression for the energy-momentum tensor for the graviton field $h_{\mu\nu}$, which in Lorentz gauge reads,
\bge
  \mathcal{T}_{\mu\nu}=\FR{c^4}{32\pi G}\la\pd_\mu h_{\rh\si}\pd_\nu h^{\rh\si}\ra.
\ede
This expression holds for general slowly varying background, so long as the propagating graviton field $h_{\mu\nu}$ has much higher frequency or much shorter wavelength than the background. Then  it is understood that the average $\la\cdots \ra$ is taken over several wavelengths in space and several periods in time. We are particularly interested in the energy density $\mathcal{T}^{00}$, which can be written in terms of TT field or in terms of polarization components as,
\bge
  \mathcal{T}^{00}=\FR{c^2}{32\pi G}\la \dot h_{ij}^{TT}\dot h_{ij}^{TT}\ra=\FR{c^2}{16\pi G}\la\dot h_+^2+\dot h_\times^2\ra,
\ede
where the difference $c^2$ in the coefficient comes from raising indices, $\pd^0=c^{-1}\di/\di t$. Now, let $V$ be a large sphere centered at the source with radius $r$ so that the boundary $\pd V$ is in the far-field region. Then, by definition, the total energy of the GW is given by $E=\int_V\di^3x\,\mathcal{T}^{00}$, and therefore the total radiation power is given by,
\bge
  P=\FR{\di E}{\di t}=\FR{\di}{\di t}\int_V\di^3x\,\mathcal{T}^{00}=c\int_V\di^3x\,\pd_i \mathcal{T}^{i0}=-c\int_{\pd V}\di\Omega\,r^2 \mathcal{T\,}^{r0},
\ede
where $\mathcal{T\,}^{r0}=\frac{c^4}{32\pi G}\la \pd^0 h_{ij}^{TT}\pd_r h_{ij}^{TT}\ra$. On the other hand, the GW in far-field region has the asymptotic form $h_{ij}^{TT}(t,r)=\frac{1}{r}f_{ij}(t-r/c)$, and therefore $\pd_r h_{ij}^{TT}=\pd^0 h_{ij}^{TT}+\order{1/r^2}$. Consequently, we have $P=-c\int\di \Omega\,r^r\mathcal{T}^{00}$, and thus,
\bge
  P=\FR{c^3r^2}{32\pi G}\int\di\Omega\,\la \dot h_{ij}^{TT}\dot h_{ij}^{TT}\ra.
\ede
For the quadrupole radiation (\ref{hTT}), we get,
\bge
  P=\FR{G}{8\pi c^5}\int\di\Omega\,\Pi_{ij,k\ell}(\hat{\mb n})\la \dddot M_{ij}\dddot M_{k\ell}\ra.
\ede
Note that the quadrupole moments are independent of direction $\hat{\mb n}$, and thus the integration over $\di\Omega$ can be carried out,
\bge
  \int\di\Omega\,\Pi_{ij,k\ell}=\FR{2\pi}{15}(11\de_{ik}\de_{j\ell}+\de_{i\ell}\de_{jk}-4\de_{ij}\de_{k\ell}),
\ede
and therefore,
\bge
\label{totalP}
  P=\FR{G}{5c^5}\la\dddot M_{ij}\dddot M_{ij}-\FR{1}{3}(\dddot M_{kk})^2\ra.
\ede

The angular momentum radiated away by the GW can be calculated similarly, but the derivation is more tedious. Here we only list several important result and the details can be found in \cite{GW}. The angular momentum, as the charge of spatial rotations, can be represented as the temporal component of the corresponding N\"other current integrated over space. For physical gravitons, this charge in TT gauge is given by,
\bge
  J^i=\FR{c^2}{32\pi G}\int\di^3x\,\Big[-\ep^{ijk}\dot h_{\ell m}^{TT}x_j \pd_k h_{\ell m}^{TT}+2\ep^{ijk}h_{\ell j}^{TT}\dot h_{\ell k}^{TT}\Big].
\ede
Then using the conservation of current and also the asymptotic form of GW in far-field region, we can find the rate of angular momentum radiated away by GWs, as,
\bge
  \FR{\di J^i}{\di t}=\FR{c^3r^2}{32\pi G}\int\di\Omega\,\Big\la-\ep^{ijk}\dot h_{\ell m}^{TT}x_j \pd_k h_{\ell m}^{TT}+2\ep^{ijk}h_{\ell j}^{TT}\dot h_{\ell k}^{TT}\Big\ra,
\ede
where the average $\la\cdots\ra$ is again taken over several periods in time or several wavelengths in space. Then substituting (\ref{hTT}) into above expression and after some algebras, we get the rate of angular momentum carried away by quadrupole radiation,
\bge
  \FR{\di J^i}{\di t}=\FR{2G}{5c^5}\ep^{ijk}\la\ddot Q_{j\ell}\dddot Q_{k\ell}\ra,
\ede
where $Q_{ij}\equiv M_{ij}-\frac{1}{3}\de_{ij}M_{kk}$ is the quadrupole mass moment.

\subsection{GWs from elliptical orbit}
\label{App_Ellip}

We generalize the discussions of circular orbits in Sec.~\ref{sec_GGW} to an elliptical orbit, for which the motion can be described again in the center-of-mass frame as,
\begin{align}
  &x =a(\cos u-e), &&y =b\sin u, &&z =0,
\end{align}
where $a$ and $b$ are semi-major axis and semi-minor axis, respectively; $e$ is the eccentricity, and the so-called eccentric anomaly $u$ is related to time $t$ via the Kepler equation,
\bge
  u-e\sin u=\omega_0 t\equiv \beta,
\ede
and $\omega_0^2\equiv Gm/a^3$. Alternatively, the orbital equation can be written in polar coordinates $(r,\psi)$ as,
\bge
  \FR{1}{r}=\FR{1+e\cos\psi}{R}.
\ede
The length scale $R$ is related to $a$ and $b$ via,
\bge
  a=\FR{R}{1-e^2}, ~~~~~~b=\FR{R}{\sqrt{1-e^2}},
\ede
and here it should be understood that $R$ is no longer the orbital radius. In polar coordinates of the orbit plane, the mass quadrupole of the binary system with an elliptical orbit is,
\bge
  M=\mu R^2\bgp \cos^2\psi & \sin\psi\cos\psi \\ \sin\psi\cos\psi & \sin^2\psi \edp.
\ede
Instead of uniform circular motion $\psi=\omega t$, the azimuthal angle $\psi(t)$ is related to the time $t$ by,
\bge
  \cos\psi=\FR{\cos u-e}{1-e\cos u},
\ede
where $u$ is the so-called eccentric anomaly, which depends on time via $\omega_0 t=u-e\sin u$ with $\omega_0^2=Gm/a^3$. As a result, the power spectrum of gravitational radiation is no longer monochromatic, and the spectrum in general has components with frequency $\omega = n\omega_0$, $n=1,2,3,\cdots$. To extract the amplitudes of these higher harmonics, we can perform Fourier expansion of the orbital coordinates. Note further that $x$ is even in $t\to -t$ and $y$ is odd, we can write,
\bge
  x (t)=\sum_{n=0}^\infty a_n \cos\omega_n t, ~~~~~y (t)=\sum_{n=1}^\infty b_n \sin\omega_n t,
\ede
and the coefficients can be obtained by inverse Fourier transform, as,
\begin{align}
  a_0=&~\FR{1}{\pi}\int_0^\pi\di\be\,x(\be)=-\FR{3}{2}ae,\\
  a_n=&~\FR{2}{\pi}\int_0^\pi\di\be\,x(\be)\cos(n\beta)=\FR{a}{n}\Big[J_{n-1}(ne)-J_{n+1}(ne)\Big],\\
  b_n=&~\FR{2}{\pi}\int_0^\pi\di\be\,y(\be)\sin(n\beta)=\FR{b}{n}\Big[J_{n-1}(ne)+J_{n+1}(ne)\Big],
\end{align}
where $n\geq 1$.

From this result we can perform a Fourier decomposition of the mass quadrupole,
\begin{align}
  M_{ab}(t)=\mu\bgp x^2 & xy \\ xy & y^2 \edp
  =\sum_{n=0}^\infty\mu R^2\bgp A_n\cos\omega_n t & C_n\sin\omega_n t \\ C_n\sin\omega_n t & B_n\cos\omega_n t \edp\equiv \sum_{n=0}^\infty M_n(t),
\end{align}
where 
\begin{align}
  A_n=&~\FR{1}{n(1-e^2)^2}\Big[J_{n-2}(ne)-J_{n+2}(ne)-2e\big(J_{n-1}(ne)-J_{n+1}(ne)\big)\Big],\\
  B_n=&~\FR{1}{n(1-e^2)}\Big[J_{n+2}(ne)+J_{n-2}(ne)\Big],\\
  C_n=&~\FR{1}{n(1-e^2)^{3/2}}\Big[J_{n+2}(ne)+J_{n-2}(ne)-e\big(J_{n+1}(ne)+J_{n-1}(ne)\big)\Big].
\end{align}
Now we can compute the power of radiation for each component $M_n(t)$ using (\ref{totalP}) and get the following result,
\bge
  P_n=\FR{32G^4\mu^2m^3}{5c^5R^5}F_n(e),
\ede
where $F_n(e)$ is defined by,
\bge
  F_n(e)=\FR{n^6(1-e^2)^9}{96}\Big[A_n^2+B_n^2+3C_n^2-A_nB_n\Big],
\ede
Then the small $e$ behavior of $F_n(e)$ can be found directly as shown in (\ref{Powern})-(\ref{Fn}).

The change of the size and the shape of the elliptical orbit due to GW radiation can be worked out by considering the energy and angular momentum $J$ carried away by GW. For angular momentum, it is given by,
\bge
  \FR{\di J^i}{\di t}=-\FR{2G}{5c^5}e^{ijk}\la \ddot{M}_{j\ell}\dddot M_{k\ell}\ra.
\ede
Putting the orbit in $(x,y)$-plane, we have $\mb J=J\hat{\mb z}$, and,
\bge
  \FR{\di J}{\di t}=\FR{4G}{5c^5}\la \ddot M_{12}(\dddot M_{11}-\dddot M_{22})\ra
  =-\FR{32}{5}\FR{G^{7/2}\mu^2 m^{5/2}}{c^5a^{7/2}}\FR{1}{(1-e^2)^2}\bigg(1+\FR{7}{8}e^2\bigg).
\ede
On the other hand, the power of GW radiation is given by,
\bge
  \FR{\di E}{\di t}=\FR{G}{5c^5}\la \dddot M_{ij}\dddot M_{ij}-\FR{1}{3}(\dddot M_{kk})^2\ra=\FR{32G^4\mu^2m^3}{5c^5a^5}\FR{1}{(1-e^2)^{7/2}}\bigg(1+\FR{73}{24}e^2+\FR{37}{96}e^4\bigg).
\ede
The energy $E$ and angular momentum $J$ is related to orbital parameters, semi-major axis $a$ and eccentricity $e$, via,
\bge
  a=\FR{Gm\mu}{2|E|},~~~~~~e^2=1+\FR{2EJ^2}{G^2m^2\mu^2}.
\ede
Therefore we get two equations for $\dot a$ and $\dot e$ as shown in (\ref{dadt}) and (\ref{dedt}).

\subsection{Sample Kozai-Lidov Solutions}

Here we present some samples of Kozai-Lidov solutions solved directly from the full 3-body equations without taking any perturbation expansion. 
 
To this end, we define the coordinate vector of $m_2$ relative to $m_1$ to be $\mb R=(x,y,z)$ and that of the tertiary body relative to the mass center of the binary to be $\mb L=(X,Y,Z)$. Therefore, the coordinate of the tertiary body in $m_1$-rest frame is given by $\wt{\mb L}\equiv \mb L+\frac{m_2}{m}\mb R$. Our sample solutions are then found by integrating the following equations written in the $m_1$-rest frame, 
\begin{align}
 \ddot{\mb R}=&-\FR{Gm_1}{|\mb R|^3}\mb R+\FR{GM}{|\wt{\mb L}-\mb R|^3}(\wt{\mb L}-\mb R)-\FR{Gm_2}{|\mb R|^3}\mb R-\FR{GM}{|\wt{\mb L}|^3}\wt{\mb L},\\
 \ddot{\wt{\mb L}}=&-\FR{Gm_1}{|\wt{\mb L}|^3}\wt{\mb L}-\FR{Gm_2}{|\wt{\mb L}-\mb R|^3}(\wt{\mb L}-\mb R)-\FR{Gm_2}{|\mb R|^3}\mb R-\FR{GM}{|\wt{\mb L}|^3}\wt{\mb L},
\end{align}
where the last two terms of each equation are from the inertial force coming from the acceleration of the $m_1$ reference frame relative to the mass center of the triple. They are just the opposite of the gravitational force per unit mass exerted on $m_1$. In the case that one body of the inner binary is light enough so that $\mu/M$ is small, the above equations of motion get simplified in the reduced coordinates of the inner binary as follows,
\begin{align}
 \ddot{\mb R}=&-\FR{Gm}{|\mb R|^3}\mb R+\FR{GM}{|{\mb L}-\mb R|^3}({\mb L}-\mb R)-\FR{GM}{|{\mb L}|^3}{\mb L},\\
 \ddot{\mb L}=&-\FR{Gm}{|{\mb L}|^3}{\mb L}-\FR{GM}{|{\mb L}|^3}{\mb L},
\end{align}
where the last term of each equation is again from the inertial force, but this time due to the gravitational force from the tertiary body alone.

\begin{figure}[tbph]
\centering
\parbox{0.4\textwidth}{\includegraphics[width=0.4\textwidth]{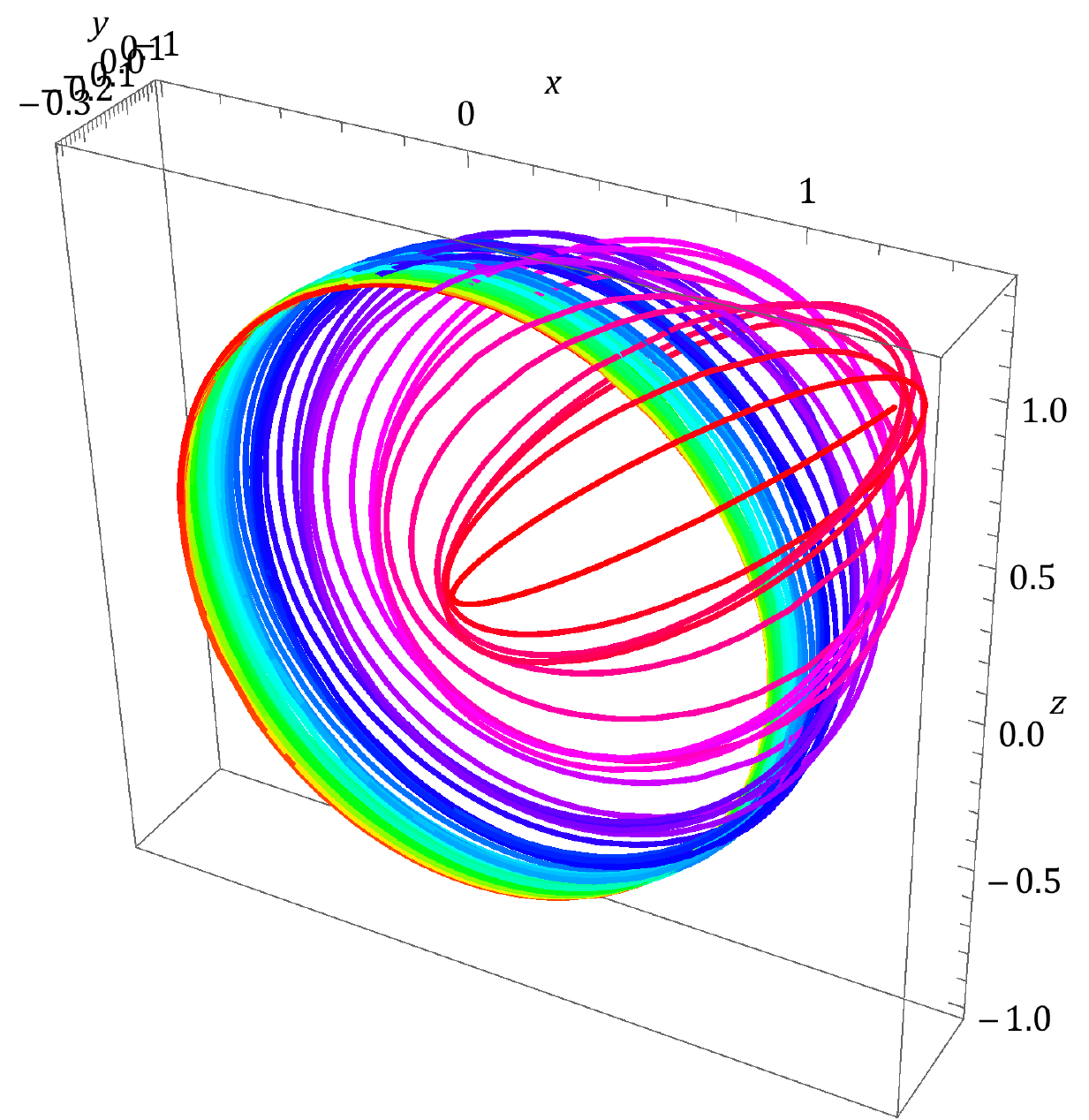}}
\hspace{1cm}
\parbox{0.4\textwidth}{\includegraphics[width=0.4\textwidth]{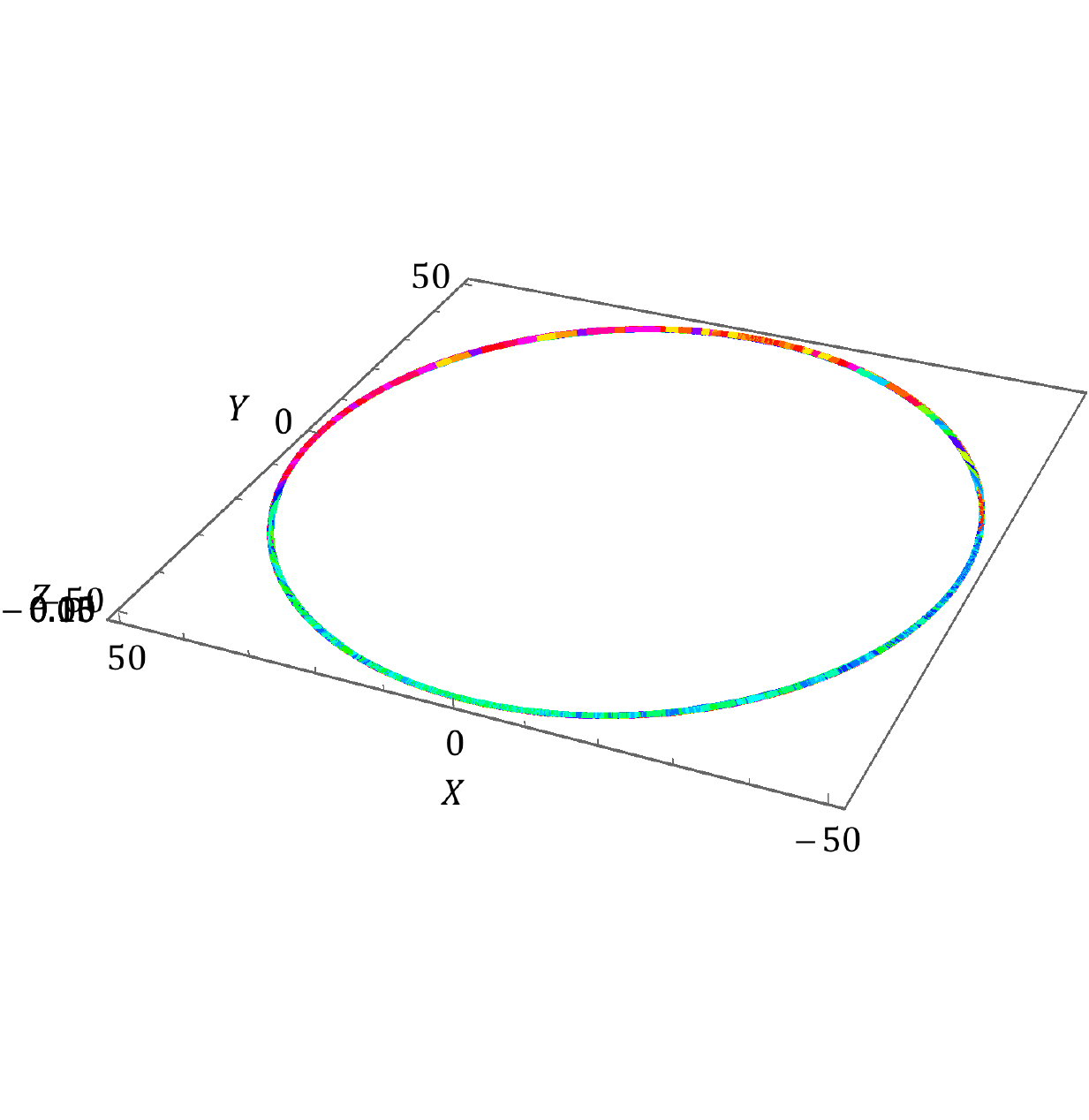}}\\[5mm]
\parbox{0.9\textwidth}{\includegraphics[width=0.9\textwidth]{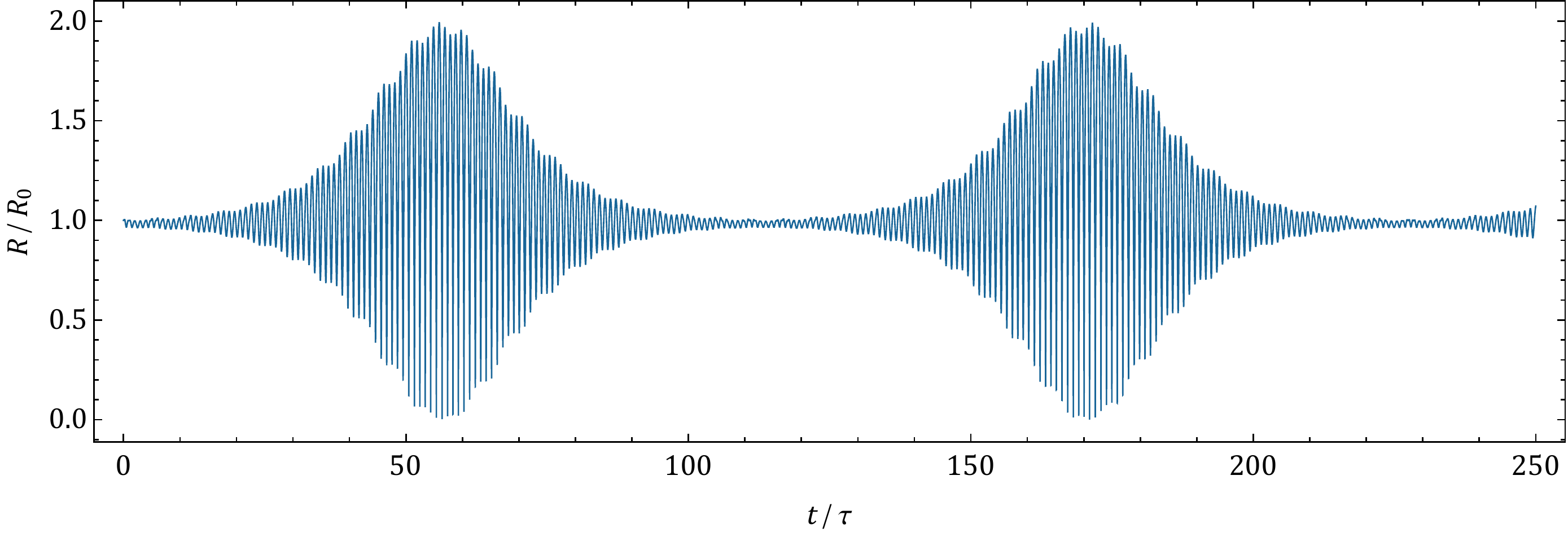}}\\[5mm]
\parbox{0.45\textwidth}{\includegraphics[width=0.45\textwidth]{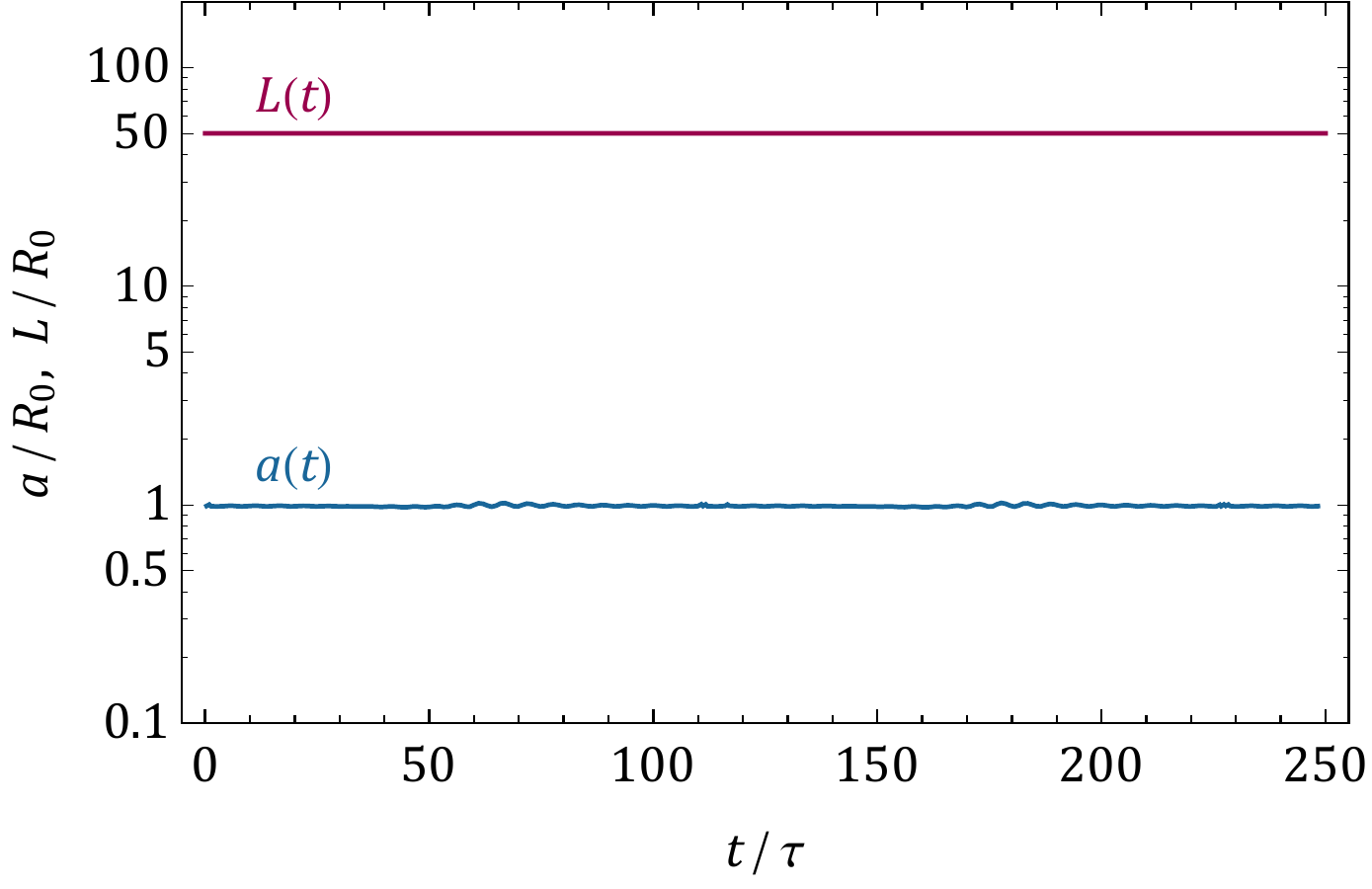}}
\parbox{0.45\textwidth}{\includegraphics[width=0.45\textwidth]{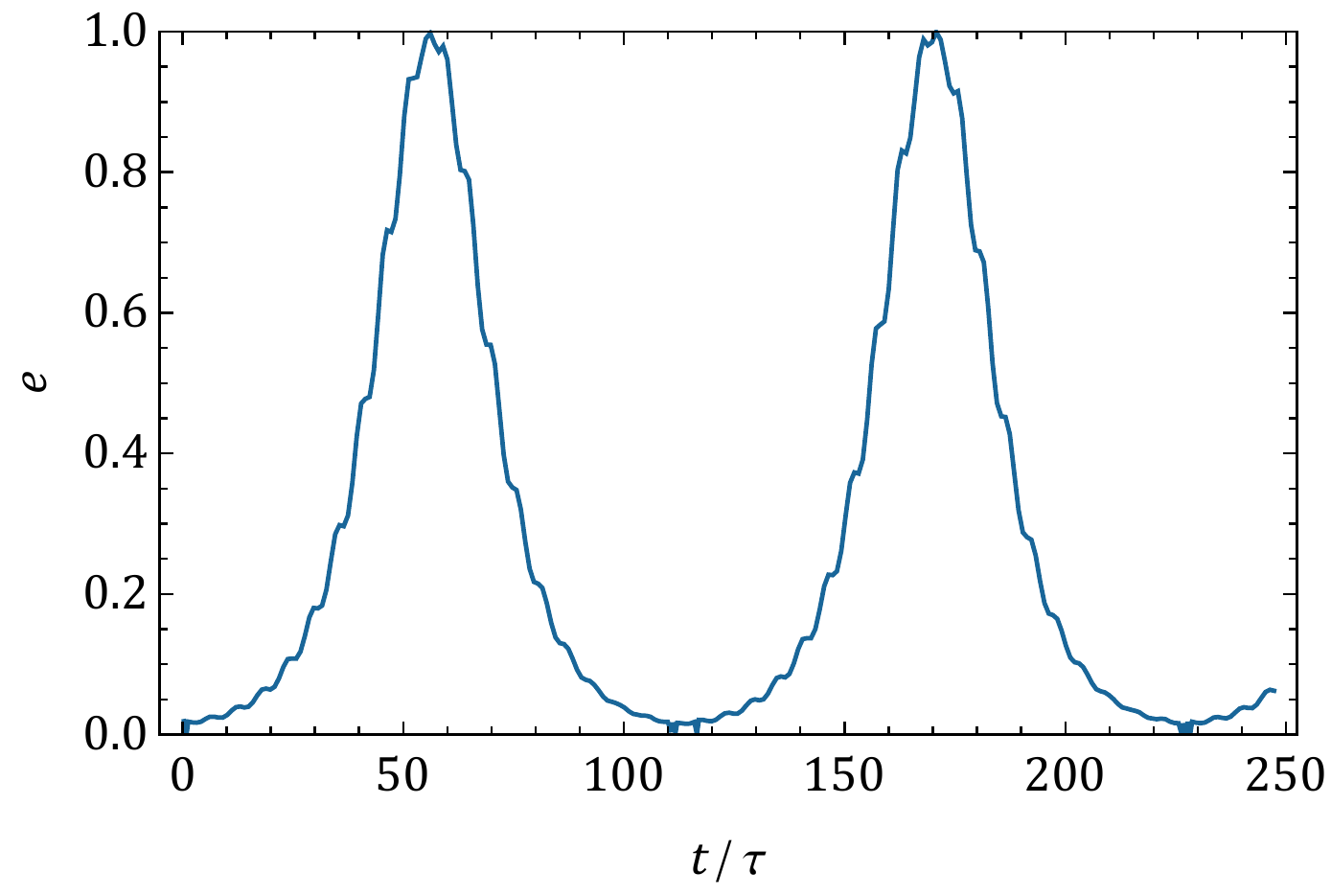}}
\caption{Sample 3d solution with $e_0=0$, $(I_0,\ga_0,\vartheta_0)=(90^\circ,0^\circ,0^\circ)$, $M/m=10^3$, and $L_0/R_0=50$. The upper-left and upper-right panels show the inner orbit and the outer orbit, respectively, in the unit of $R_0$. The second row shows the binary separation as a function of time. The lower-left and lower-right panels show $a(t)$ and $L(t)$, and $e(t)$ respectively. The time $t$ is in the unit of inner orbital period $\tau=2\pi\omega^{-1}$.}
\label{3dsol1}
\end{figure}
\begin{figure}[tbph]
\centering
\parbox{0.45\textwidth}{\includegraphics[width=0.45\textwidth]{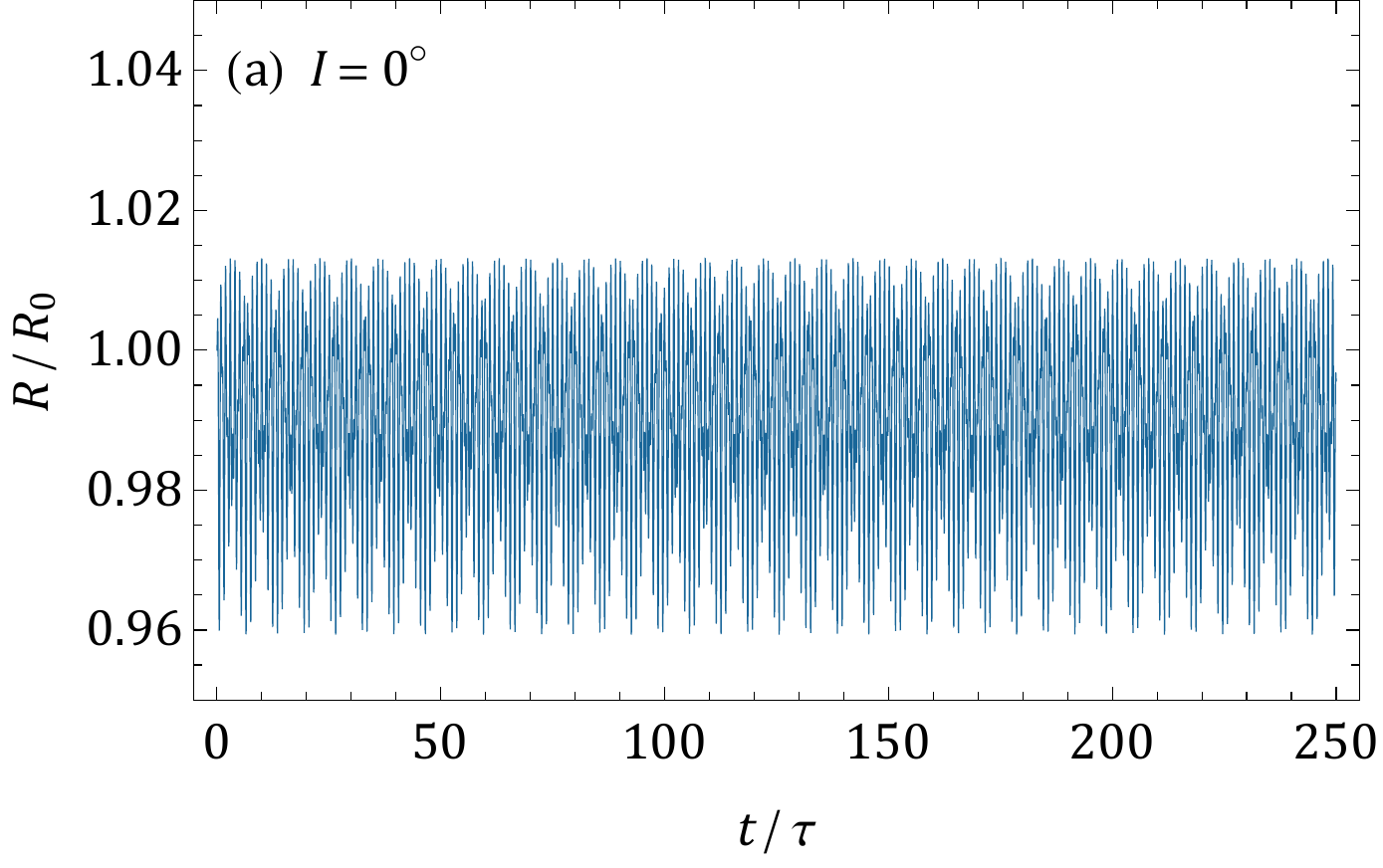}}
\parbox{0.45\textwidth}{\includegraphics[width=0.45\textwidth]{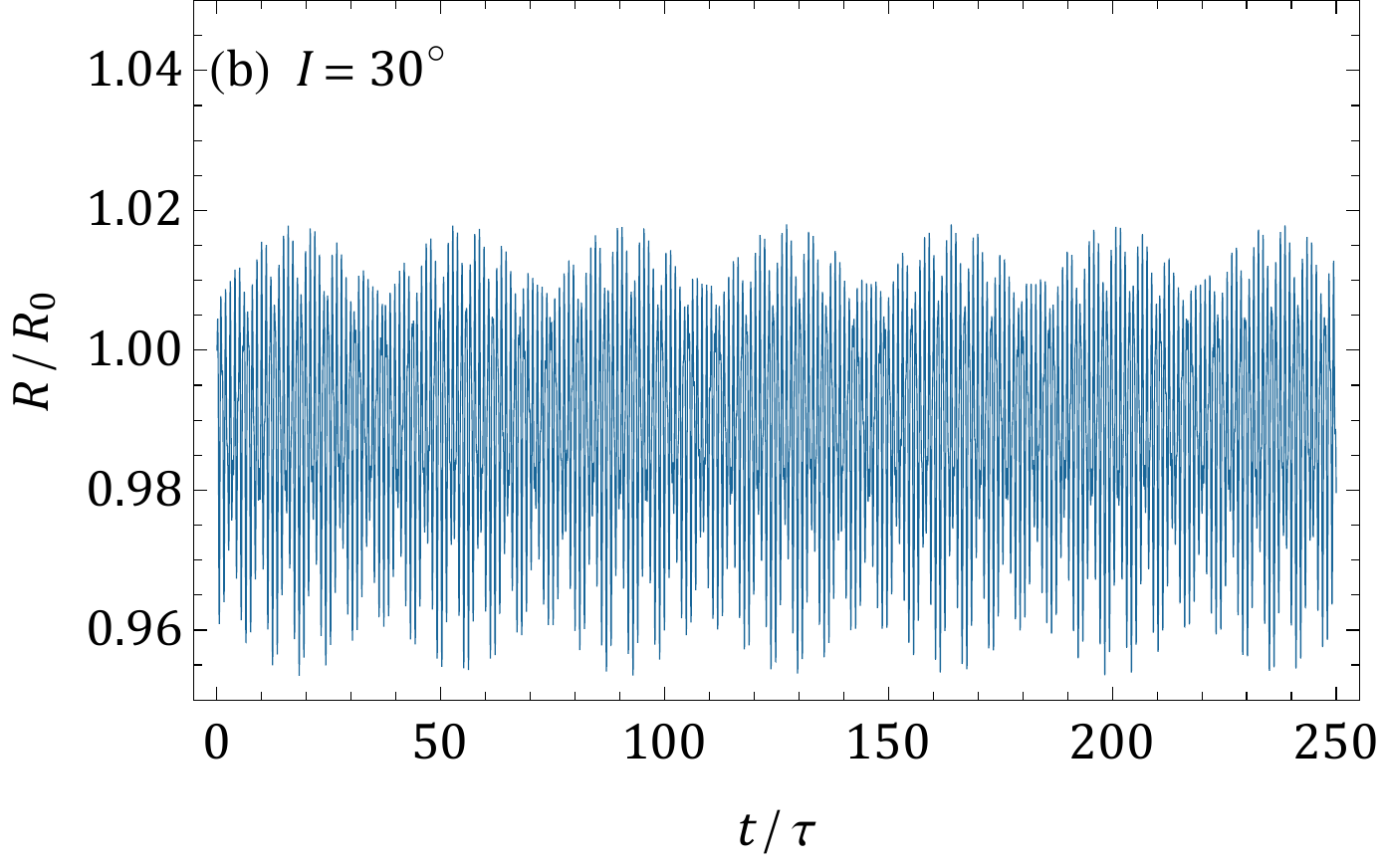}}\\
\parbox{0.45\textwidth}{\includegraphics[width=0.45\textwidth]{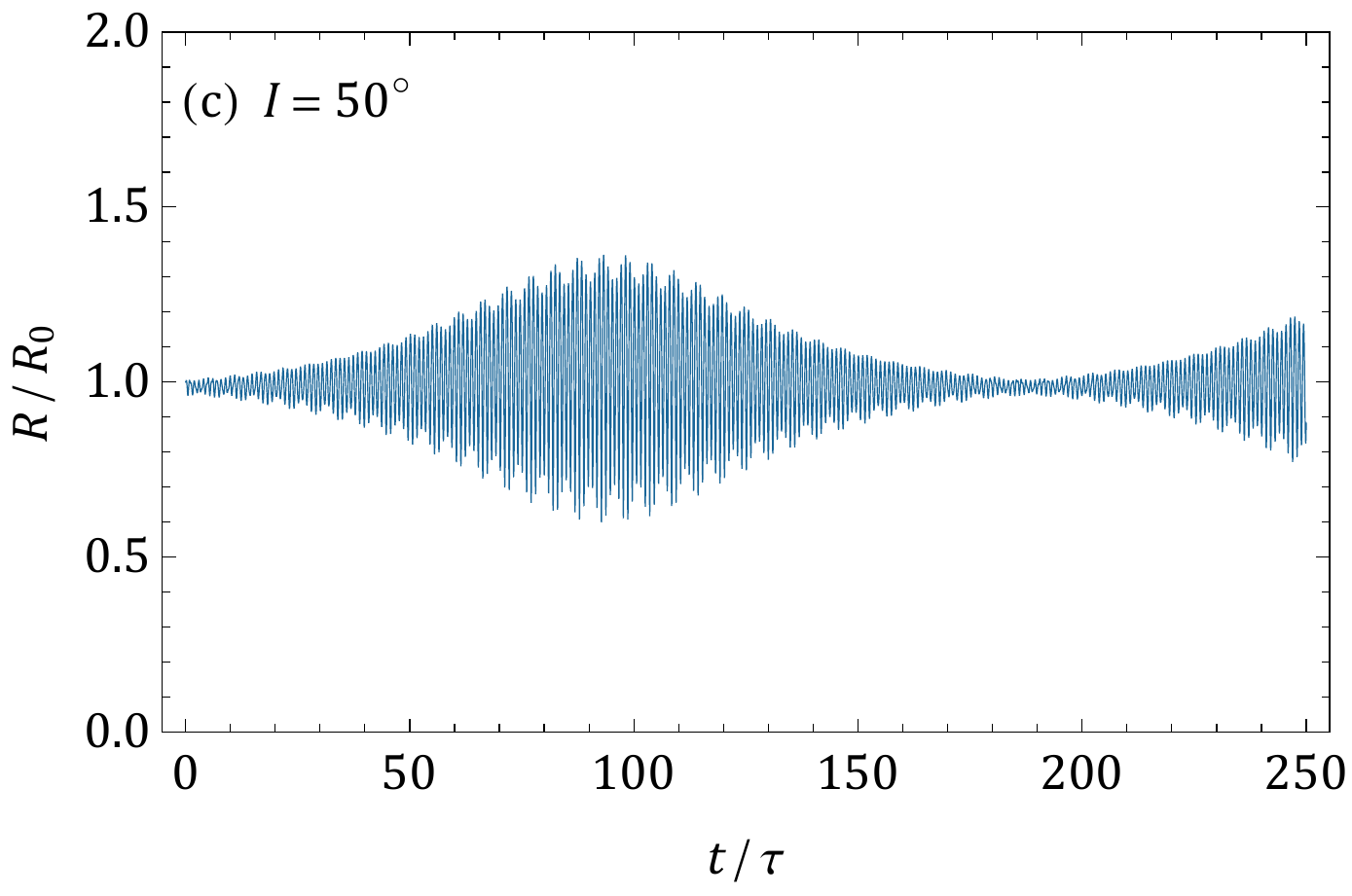}}
\parbox{0.45\textwidth}{\includegraphics[width=0.45\textwidth]{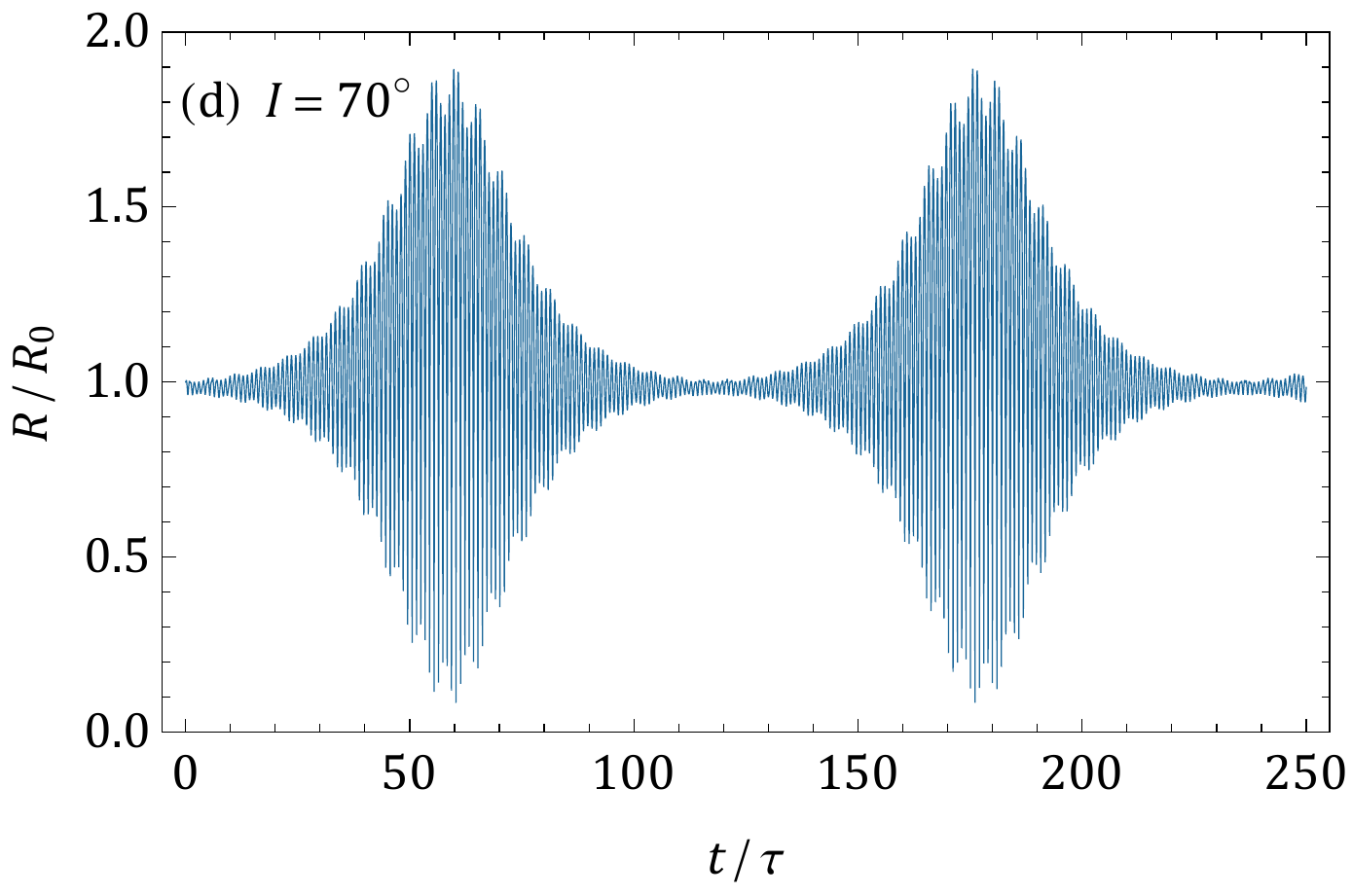}}
\caption{$R(t)$ with initial inclination $I_0=0^\circ$ (upper left), $30^\circ$ (upper right), $50^\circ$ (lower left), and $70^\circ$ (lower right), respectively. All other parameters are taken to be the same as in Fig.~\ref{3dsol1}.}
\label{3dsolR}
\end{figure}

We choose the initial conditions such that the tertiary body is initially in a circular orbit relative to the mass center of the binary,
\bge
  (X_0,Y_0,Z_0)=(L_0,0,0),~~~~~~~~~(\dot X_0,\dot Y_0,\dot Z_0)=(0,\sqrt{\FR{G(M+m)}{L_0}},0),
\ede
and that the binary system has initial eccentricity $e_0$ in a plane with inclination $I$, the longitude of ascending node $\vartheta$, and the angle from the ascending node to the periastron, or the argument of the periastron\footnote{It is customary to use $g_1$ to denote the argument of the periastron, but we choose to use $\ga$ here to avoid confusion with the $g(e)$ function defined in (\ref{ge}).} $\ga$, i.e.,
\begin{align}
  x_0=&~R_0(1-e_0)(\cos\ga\cos\vartheta-\sin\ga\cos I\sin\vartheta),\\
  y_0=&~R_0(1-e_0)(\cos\ga\sin\vartheta+\sin\ga\cos I\cos\vartheta),\\
  z_0=&~R_0(1-e_0)\sin\ga\sin I,\\
  \dot x_0=&~\sqrt{\FR{1+e_0}{1-e_0}\FR{Gm}{R_0}}(-\sin\ga\cos\vartheta-\cos\ga\cos I\sin\vartheta),\\
  \dot y_0=&~\sqrt{\FR{1+e_0}{1-e_0}\FR{Gm}{R_0}}(-\sin\ga\sin\vartheta+\cos\ga\cos I\cos\vartheta),\\
  \dot z_0=&~\sqrt{\FR{1+e_0}{1-e_0}\FR{Gm}{R_0}}\sin\ga\sin I.
\end{align}

With the above initial conditions, we can solve the equations of motion numerically. We present several samples. First, we choose $e_0=0$ and $(I_0,\ga_0,\vartheta_0)=(90^\circ,0^\circ,0^\circ)$. We also choose $M/m=10^3$ and $L_0/R_0=50$. In Fig.\;\ref{3dsol1}, we plot the 3d diagrams of inner orbit (upper left), large orbit (upper right), the function $R(t)$ (second row), $a(t)$ and $L(t)$ (lower left), and $e(t)$ (lower right), respectively. The time $t$ is in the unit of inner orbital period $\tau=2\pi\omega^{-1}$.

We can observe 2 Kozai cycles from both $R(t)$ and $e(t)$, while both $a(t)$ and $L(t)$ are approximately conserved. It's also easy to see that the Kozai time scale is much greater than the period of the large orbit. (In this example, the period of the large orbit $2\pi\Omega^{-1}\simeq  11.2\tau$.) The generation of eccentricity can be seen from both the 3d plot and the behavior of $R(t)$. In addition, $L(t)$ has very weak time dependence and the large orbit remains essentially circular.

In Fig.\;\ref{3dsolR}, we present $R(t)$ for several different choices of initial inclination $I_0=0^\circ,\;30^\circ,\;50^\circ,\;70^\circ$, respectively, while all other initial conditions remain unchanged. We can see from the behavior of $R(t)$ that the mechanism of eccentricity generation works effectively in our examples only for $I_0=50^\circ$ and $I_0=70^\circ$, as well as the above solution $I_0=90^\circ$. This confirms the Kozai-Lidov calculation of the critical inclination $I_c=\arccos\sqrt{3/5}\simeq 39^\circ$.

\subsection{Numerical Memo}
\label{App_Memo}

\begin{figure}[tbph]
\centering
\includegraphics[width=0.48\textwidth]{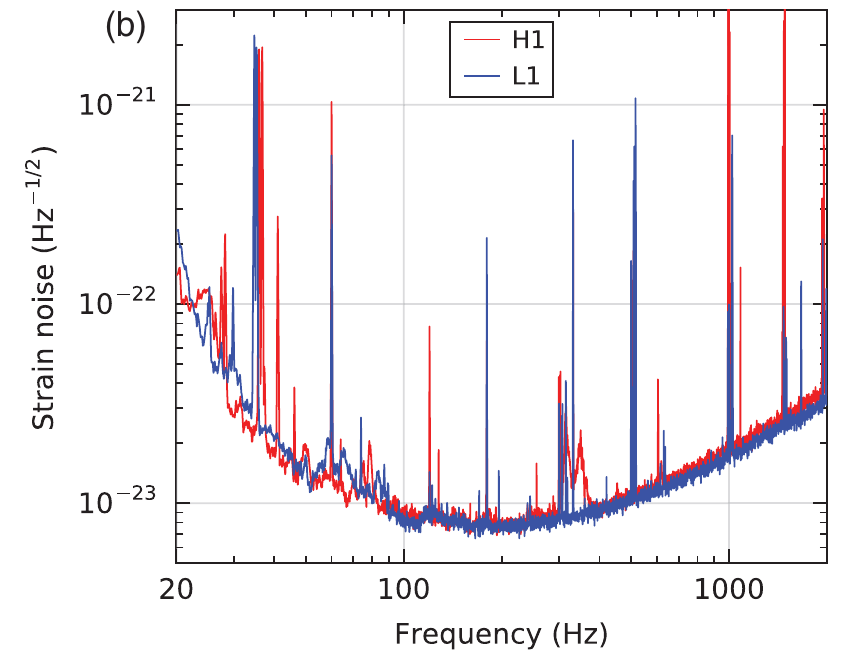}
\includegraphics[width=0.48\textwidth]{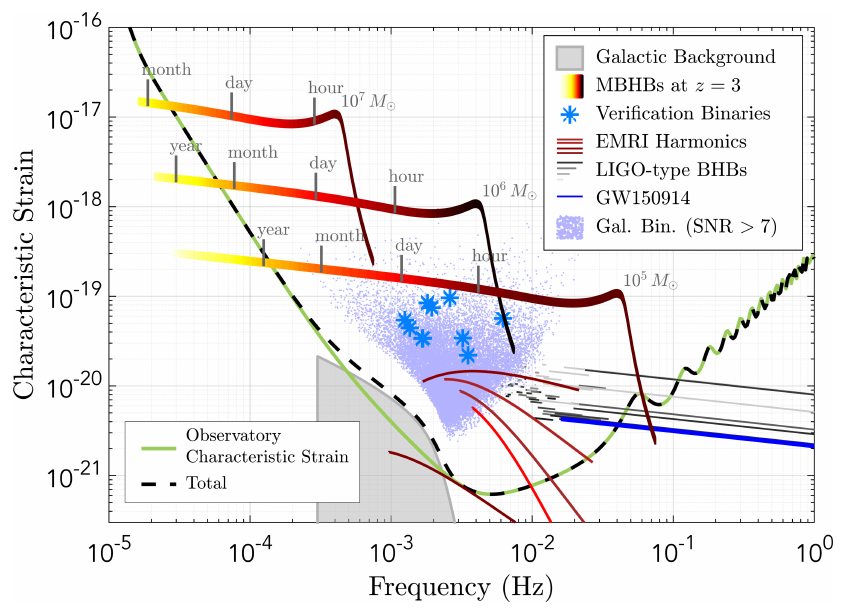}
\caption{Sensitivities of LIGO \cite{Abbott:2016blz} (left) and LISA (right) experiments.}
\label{fig_noise}
\end{figure}

In this appendix we list some numbers related to binary systems that are potentially observable to LIGO/LISA. Firstly we quote plots of the sensitivity ranges of both experiments in Fig.\;\ref{fig_noise}.

The LIGO experiment can observe binary systems with mass $\order{M_\odot}\sim\order{10M_\odot}$ or even $\order{10^2M_\odot}$  if they exist, while LISA is planned to be optimized for binary system of mass $\order{10^{5}M_\odot}\sim \order{10^7M_\odot}$.

\begin{figure}[tbph]
\centering
\includegraphics[width=0.45\textwidth]{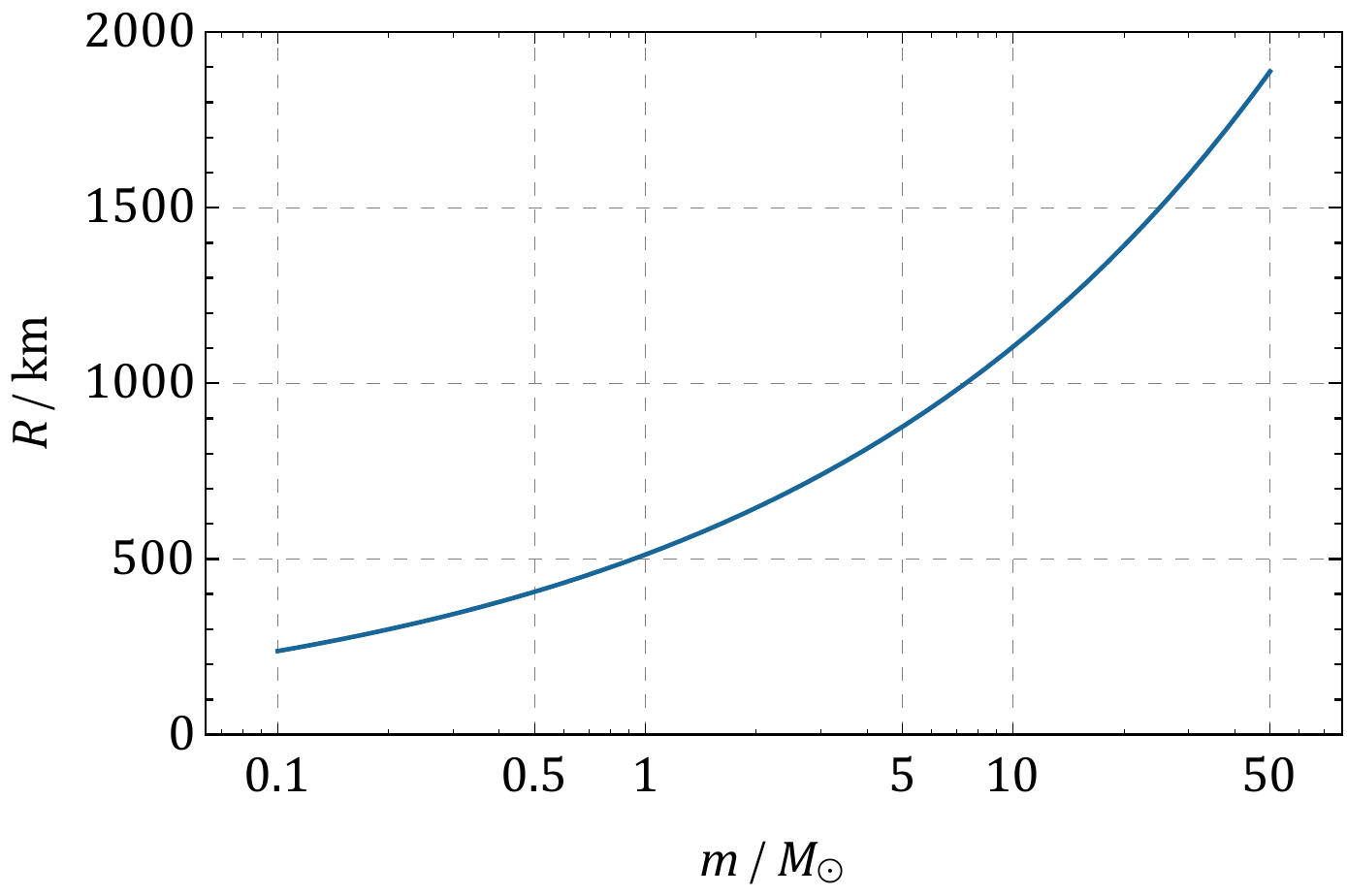}\hspace{3mm}
\includegraphics[width=0.45\textwidth]{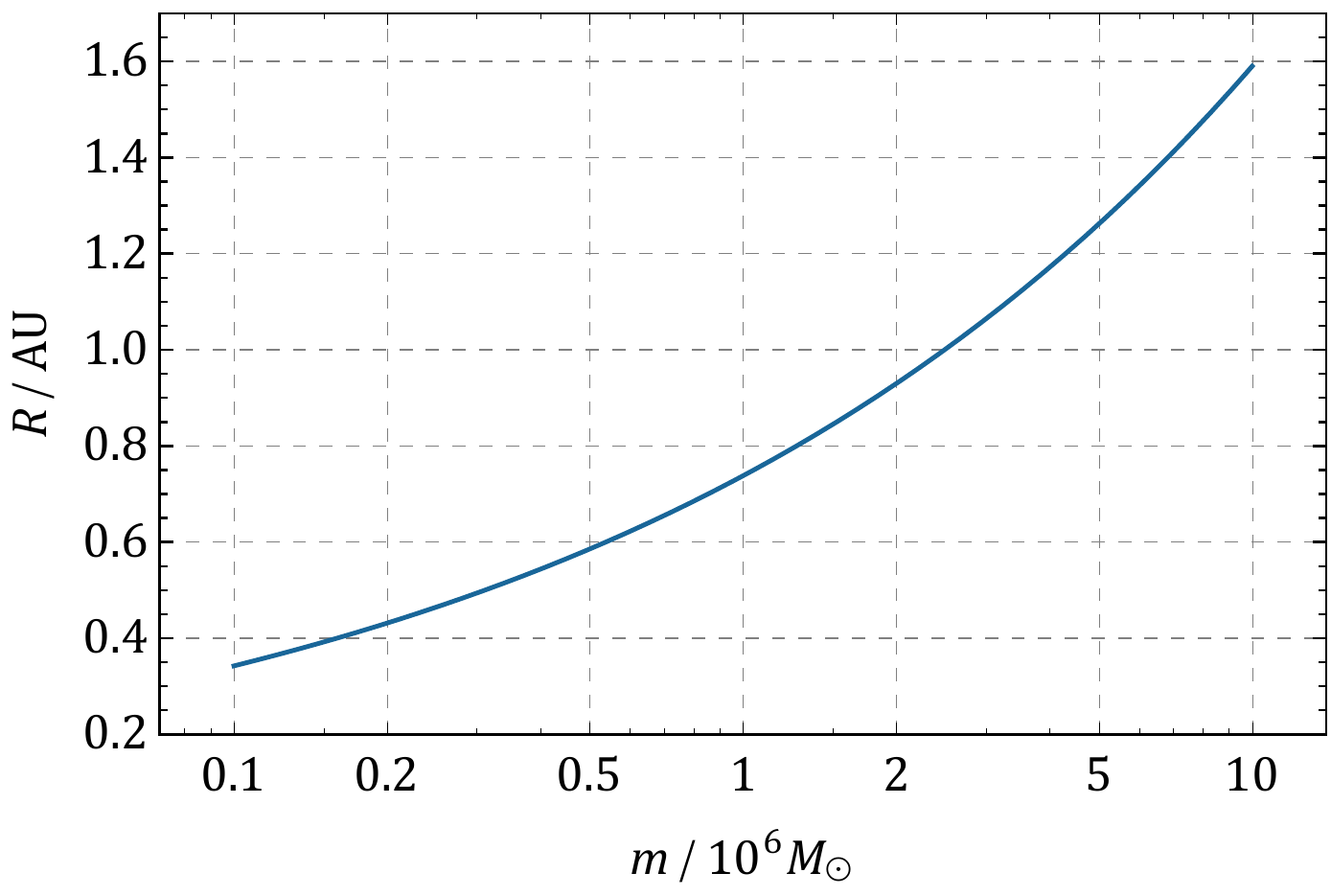} 
\caption{The orbital radius of a binary system of mass $m$ when entering the sensitivity window of LIGO (left) and LISA (right).}
\label{fig_radius}
\end{figure}

Since we are interested in the background perturbations to inspiraling binaries, it is the low frequency part that will be most relevant. For an order-of-magnitude estimate, we can take the lowest sensitive frequency of LIGO (LISA) to be 10Hz ($10^{-4}$Hz). Given this minimal frequency $\omega_\text{min}$, we are interested in the orbital radius of the binary system $R_\text{max}=R_\text{max}(m)$ as a function of its total mass $m$ when it enters the sensitivity band of both experiments, which is given by,
\bge
\label{Rmax}
  R_\text{max}(m)=\bigg(\FR{Gm}{\omega_\text{min}^2}\bigg)^{1/3}.
\ede
We plot this function for both experiments in Fig.\;\ref{fig_radius}.

\end{appendix}

%\end{fmffile}
\end{document}